\documentclass[acmtog]{acmart}
\AtBeginDocument{%
  }

\setcopyright{cc}
\setcctype{by}
\acmJournal{TOG}
\acmYear{2025} \acmVolume{44} \acmNumber{1} \acmArticle{10} \acmMonth{2} \acmDOI{10.1145/3711853}

\usepackage[newfloat]{minted} 
\usepackage{listings}

\usepackage{tcolorbox}
\usepackage{multirow}
\usepackage{listings}

\usepackage{mdframed}
\usepackage{setspace}
\usepackage{bm}
\usepackage{ulem}
\definecolor{hlcol}{RGB}{255,200,200}
\newminted{cpp}{autogobble,escapeinside=||,fontsize=\footnotesize}
\newminted{python}{autogobble,escapeinside=||,fontsize=\footnotesize}

\BeforeBeginEnvironment{minted}{
  \begin{tcolorbox}[
    arc=1pt,
    colback=black!3!white,
    colframe=black!16!white,
    boxrule=0.3mm,
    before skip=1mm,
    after skip=1mm,
    top=0.5mm,
    bottom=0.5mm,
    left=1mm,
    right=1mm,
  ]
}
\AfterEndEnvironment{minted}{
  \end{tcolorbox}
}
\AtBeginEnvironment{minted}{\setstretch{1.2}}

\newenvironment{code}{\captionsetup{type=listing}}{}
\SetupFloatingEnvironment{listing}{name=Listing}
\usepackage{url}

\DeclareUrlCommand\customurl{%
  \renewcommand\UrlLeft{\uline\bgroup}%
  \renewcommand\UrlRight{\egroup}}

\newcommand{\revision}[1]{\textcolor{black}{#1}}
\newcommand{\revisionB}[1]{\textcolor{black}{#1}}

\definecolor{lightred}{rgb}{1,0.4,0.4}
\definecolor{lightyellow}{rgb}{1,1,0.4}
\definecolor{lightorange}{rgb}{1,0.647,0.4}

\newcommand{\dir}{\omega}
\newcommand{\point}[1]{\mathbf{#1}}
\newcommand{\px}{\point{x}}
\newcommand{\wi}{\dir'}
\newcommand{\wo}{\dir}
\newcommand{\rarrow}{\rightarrow}

\newcommand{\sRad}{L}
\newcommand{\sRadE}{Q}
\newcommand{\sRedRadE}{\widebar{\sRadE}}
\newcommand{\sRadI}{\sRad}
\newcommand{\sRadS}{S}
\newcommand{\sRadO}{\sRad_o}
\newcommand{\sRadOI}{\sRad_{o,i}}

\newcommand{\sCoeff}{\mu}
\newcommand{\sExt}{\sCoeff_t}
\newcommand{\sScatt}{\sCoeff_s}
\newcommand{\sAbs}{\sCoeff_a}
\newcommand{\sPF}{f_\text{p}}
\newcommand{\sAlbedo}{\alpha}

\newcommand{\sTrans}{\textrm{T}}
\newcommand{\sOptDepth}{\tau}

\newcommand{\sDensity}{\rho}
\newcommand{\sCross}{\sigma}

\newcommand{\sPrim}{\mathcal{P}}
\newcommand{\sSegment}{\mathcal{S}}
\newcommand{\sSegmentSet}{S}
\newcommand{\stsegment}{\widehat{t}}

\newcommand{\sKernel}{K}

\newcommand{\ray}{\mathrm{r}}

\newcommand{\Sphere}{{\mathcal{S}^2}}
\newcommand{\diff}{\text{d}}

\newcommand{\sPDF}{p}

\DeclareMathOperator\erf{erf}

\begin{document}
\title{Don't Splat your Gaussians: Volumetric Ray-Traced Primitives for Modeling and Rendering Scattering and Emissive Media}

\author{Jorge Condor}
\email{jorge.condor@usi.ch}
\orcid{0000-0002-9958-0118}
\affiliation{%
  \institution{Meta Reality Labs}
  \streetaddress{Giesshübelstrasse 30}
  \city{Zurich}
  \state{Zurich}
  \country{Switzerland}
  \postcode{8045}
}%
\affiliation{%
  \institution{USI Lugano}
  \streetaddress{Via Giuseppe Buffi 13}
  \city{Lugano}
  \state{Ticino}
  \country{Switzerland}
  \postcode{6900}
}%
\author{Sébastien Speierer}
\email{speierers@meta.com}
\orcid{0000-0001-6919-7567}
\affiliation{%
  \institution{Meta Reality Labs}
  \streetaddress{Giesshübelstrasse 30}
  \city{Zurich}
  \country{Switzerland}
  \postcode{8045}
}%
\author{Lukas Bode}
\email{lbode@meta.com}
\orcid{0000-0002-8710-8561}
\affiliation{%
  \institution{Meta Reality Labs}
  \streetaddress{Giesshübelstrasse 30}
  \city{Zurich}
  \country{Switzerland}
  \postcode{8045}
}%
\author{Aljaž Božič}
\email{aljaz@meta.com}
\orcid{0009-0002-2985-6921}
\affiliation{%
  \institution{Meta Reality Labs}
  \streetaddress{Giesshübelstrasse 30}
  \city{Zurich}
  \country{Switzerland}
  \postcode{8045}
}%
\author{Simon Green}
\email{simongreen@meta.com}
\orcid{0009-0006-5323-1170}
\affiliation{%
  \institution{Meta Reality Labs}
  \city{London}
  \country{United Kingdom}
  \postcode{8045}
}%
\author{Piotr Didyk}
\email{piotr.didyk@usi.ch}
\orcid{0000-0003-0768-8939}
\affiliation{%
  \institution{USI Lugano}
  \streetaddress{Via Giuseppe Buffi 13}
  \city{Lugano}
  \country{Switzerland}
  \postcode{6900}
}%
\author{Adrián Jarabo}
\email{ajarabo@meta.com}
\orcid{0000-0001-9000-0466}
\affiliation{%
  \institution{Meta Reality Labs}
  \streetaddress{}
  \city{Zaragoza}
  \country{Spain}
}%
\renewcommand{\shortauthors}{Condor, et al.}
\begin{teaserfigure}
  \centering
  \includegraphics[width=6.9in]{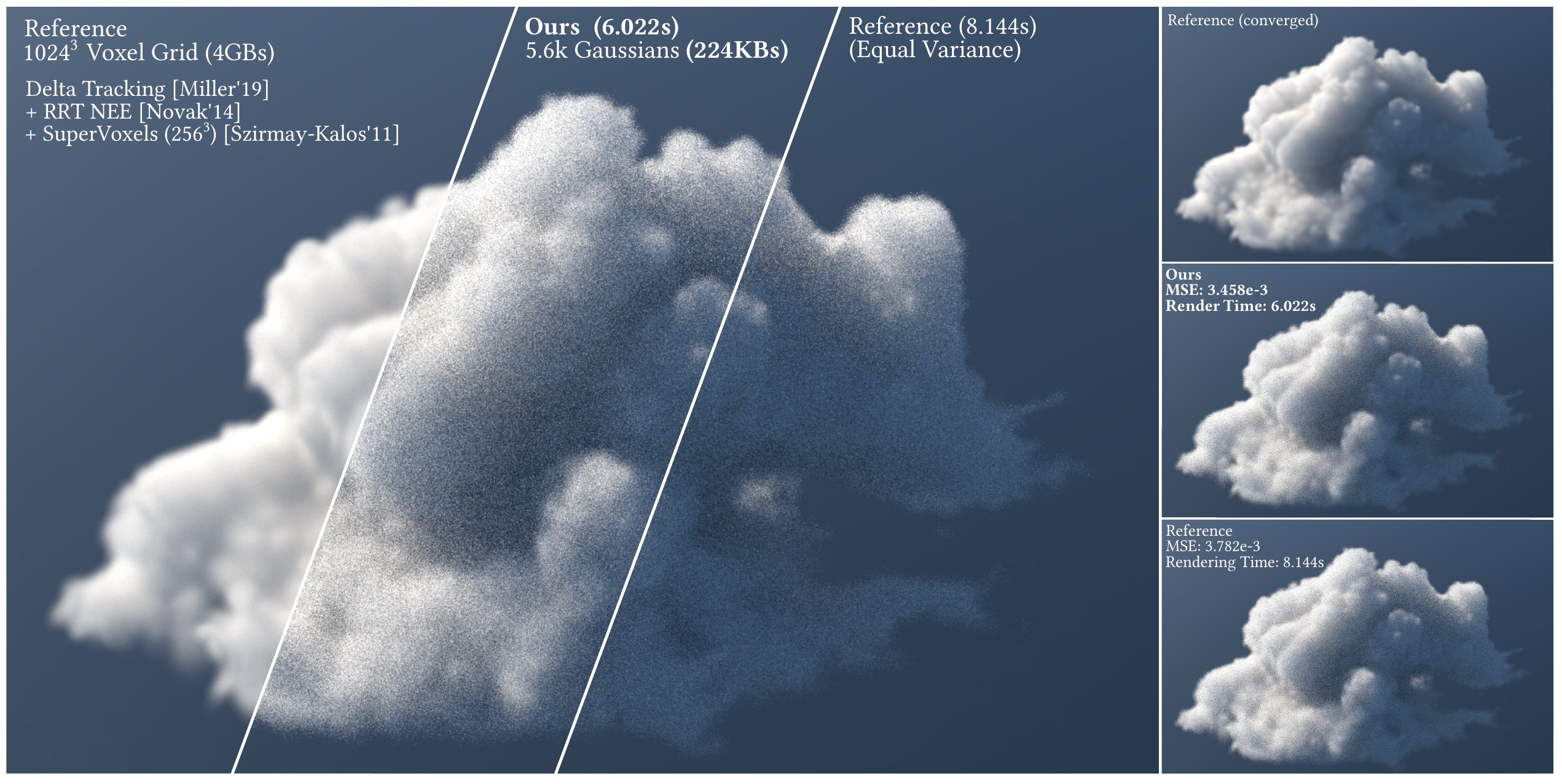}
  \caption{
  We represent a complex volumetric cloud using traditional grid-based methods (left and right, $1024^3$ voxel grid resolution, 4GBs) and our primitives-based representation using Gaussian kernels (middle, 5.6k primitives, 224KBs), and render it with volumetric path tracing. Our method achieves substantial speedups thanks to the analytical transmission estimation and sampling, our efficient rendering approach and its extremely compact representation. When compared to the original asset, at a potential cost of detail (Figure~\ref{fig:gaussian_converged_comp}), we provide large performance and memory compression benefits. Asset is part of the Walt Disney Animation Studios \emph{cloud} dataset (CC-BY-SA 3.0). Rendering times reported on a NVIDIA A6000.}
  \label{fig:teaser}
  \Description[Teaser showcasing the Disney Cloud]{The Disney cloud rendered with our method vs a grid-based alternative. We show better compression and rendering times}
\end{teaserfigure}
\begin{abstract}
Efficient scene representations are essential for many computer graphics applications. A general unified representation that can handle both surfaces and volumes simultaneously, remains a research challenge. 
In this work we propose a compact and efficient alternative to existing volumetric representations for rendering such as voxel grids. Inspired by recent methods for scene reconstruction that leverage mixtures of 3D Gaussians to model radiance fields, we formalize and generalize the modeling of scattering and emissive media using mixtures of simple kernel-based volumetric primitives. We introduce closed-form solutions for transmittance and free-flight distance sampling for different kernels, and propose several optimizations to use our method efficiently within any off-the-shelf volumetric path tracer.
We demonstrate our method in both forward and inverse rendering of complex scattering media. Furthermore, we adapt and showcase our method in radiance field optimization and rendering, providing additional flexibility compared to current state of the art given its ray-tracing formulation. We also introduce the Epanechnikov kernel and demonstrate its potential as an efficient alternative to the traditionally-used Gaussian kernel in scene reconstruction tasks. The versatility and physically-based nature of our approach allows us to go beyond radiance fields and bring to kernel-based modeling and rendering any path-tracing enabled functionality such as scattering, relighting and complex camera models.
Code is available at \customurl{https://github.com/facebookresearch/volumetric_primitives}
\label{sec:abstract}
\end{abstract}



\received{17 July 2024}
\received[revised]{1 November 2024}
\received[accepted]{2 December 2024}

\maketitle

\section{Introduction}
\label{sec:intro}

Volumetric representations of appearance have found good success at representing complex appearances such as cloth and hair~\cite{shroder11, zhao11, khungurn2015, aliaga17}, trees~\cite{neyret98,loubet2017}, clouds and smoke~\cite{kallweit2017}, or particle aggregates~\cite{moon07, meng15, mueller16efficient}, to name a few. 
In this work we propose a compact volumetric representation of appearance, that allows efficient physics-based rendering of both scattering and emissive media.

Most previous approaches for modeling volumetric appearances leverage grid-like representations of appearance such as voxels-grids, potentially with a multilevel underlying hierarchy, for which highly efficiently libraries exist~\cite{museth2013vdb}. 
Voxel grids are standard in rendering, and used in most research~\cite{jakob2010mitsuba, jakob2022mitsuba3, pbrt} and production renderers~\cite{fajardo18arnold, renderman, hyperion, manuka}.
Voxel-grids are flexible, provide $O(1)$ query time, and spatial connectivity for e.g., downsampling and level-of-detail representations~\cite{heitz2015sggx, loubet2017}. Unfortunately, they scale poorly in terms of memory, specially when representing very fine detail and sparse structures. While hierarchical grid-based methods~\cite{ProductionVolumeRendering2017} have proven very effective at subdividing complex heterogeneous models into simpler local components, their building blocks are still grid-based, essentially inheriting the same problems. For an efficient integration of e.g., transmittance, voxel grids require the use of stochastic tracking techniques~\cite{novak2018monte}, which introduces additional variance when rendering media. 

In parallel, volume rendering techniques and continuous volumetric representations have recently seen unprecedented interest in the fields of computer vision and image-based graphics, spearheaded by efforts such as neural radiance fields~\cite{mildenhall2020}. These methods allow to capture and render photorealistic three-dimensional scenes, by optimizing an underlying volumetric representation of matter. Different volumetric representations have been proposed, including point-based methods~\cite{kerbl20233Dgaussians, sainz04}, implicit neural models~\cite{mildenhall2020, mipnerf, barron2022mipnerf360}, explicit voxel-based representations~\cite{yu21plenoxels, yu2021plenoctrees}, and hybrid approaches~\cite{mueller2022instant, pointnerf, Lombardi21}. 
While implicit neural models generally achieve the highest compression rates, point-based methods can be considered as the state of the art in terms of image quality and rendering speed for radiance fields rendering, while still being substantially more compact than voxel-grid structures. However, while their derivation starts from physics (i.e., the volume rendering equation), these methods in general are not suitable for their use within a physics-based renderer, given the fairly large assumptions and simplifications imposed in their image formation model to favour speed and ease of optimization. 

Inspired by the success of recent 3D Gaussian-based representations of radiance fields~\cite{kerbl20233Dgaussians}, we propose a new representation for general scattering and emissive media based on mixtures of three-dimensional kernel-based volumetric primitives. Each primitive statistically represents a spatial distribution of matter with the same optical properties. We introduce this representation inside the radiative transfer theory (RTT)~\cite{chandrasekhar1960radiative} and derive closed-form solutions for transmittance and emission without the need of stochastic sampling, as well as sampling routines for computing inscattering. 

While the scene representation is similar to the Gaussian splatting technique~\cite{kerbl20233Dgaussians}, the image formation model is fundamentally different: Kerbl et al. assume that each primitive becomes a billboard oriented towards the camera and therefore the density along the ray becomes a sum of delta functions; this allows them to implement their model in an extremely efficient rasterization pipeline. In contrast, our radiative primitive-based formulation preserves the primitive's three-dimensional density and integrates following the physical transport process. This is crucial for physics-based rendering since it allows preserving reciprocity, while solving some view-dependent primitive ordering problems in Gaussian splatting. 
Our theoretical framework is general and supports a wide range of three-dimensional kernels. We demonstrate our work using the Gaussian kernel employed by Kerbl et al.~\shortcite{kerbl20233Dgaussians}, but also on 3D Epanechnikov kernels, implementing closed-form solutions for volumetric transmittance, emission, and free-flight sampling.
We implement our work by leveraging hardware-accelerated ray tracing for finding the primitives' bounds and integrating their contribution along the ray. This fits well inside modern ray-tracing-based renderers, allowing us to compute transmittance and emission for both primary and secondary rays, needed for multiple scattering. We also derive the adjoint of our image formation model, which we demonstrate in inverse rendering applications.  
We showcase our primitive-based volumetric model through several applications both in forward and inverse rendering: 1) traditional forward rendering of scattering media, where we demonstrate improved performance over voxel-grids using local aggregated statistics and state of the art transmittance estimators; 2) inverse rendering of scattering and purely absorptive media; 3) inverse tomographic reconstructions from focus stacks using telecentric cameras and 4) radiance field optimization and rendering of complex scenes (both real and synthetic), with increased generality and controllability. Furthermore, our integration into the general RTT allows for future extensions in relighting and scattering.
In short, our \textbf{contributions} are:
\begin{itemize}
    \item a novel kernel-based primitive representation for volumetric media that fits into the radiative transfer framework and can be integrated in any physics-based rendering engine;
    \item closed-form expressions for transmittance and emission, as well as distance sampling routines proportional to transmittance, for 3D Gaussian and Epanechnikov kernels;
    \item an efficient ray tracing-based implementation for solving light transport using our novel volumetric representation for both scattering and emissive (radiance field) media;
    \item and adjoint derivatives of our forward methods for solving inverse reconstruction problems efficiently.%
\end{itemize}

\section{Related Work}
\label{sec:relatedwork}

\paragraph*{Volumetric light transport}
Simulating light transport in media has been thoroughly studied in computer graphics~\cite{novak2018monte}. It involves solving the radiative transfer equation (RTE)~\cite{chandrasekhar1960radiative}, which has been extended to account for anisotropic \cite{jakob2010radiative}, refractive~\cite{ament14refractive}, or spatially correlated \cite{jarabo2018radiative, bitterli2018radiative} media. These generalizations make it possible to render a wider range of light transport phenomena. 

Numerical estimation of transmittance is at the core of modern volumetric light transport simulation \cite{novak2018monte}; via biased quadrature rules (ray-marching)~\cite{tuy1984direct,munoz2014higher} and modern unbiased variants~\cite{kettunen21}, to stochastic tracking algorithms (null-scattering estimators)~\cite{woodcock}, including extensions with variance reduction via control variates~\cite{novak14,szirmay2017unbiased,crespo2021primary}, alternative formulations based on power-series~\cite{georgiev19}, or differentiable formulations targeting inverse rendering~\cite{nimierdavid2022unbiased}. All these approaches require tight estimates of the maximum density (\emph{majorant}) for performance, either obtained in precomputation~\cite{yue2010unbiased,szirmay2011free} or estimated on-the-fly~\cite{carter1972monte, kutz2017spectral, galtier2013integral, misso2023progressive}. Our approach, on the other hand, allows to compute transmittance in closed-form, as the product of transmittance of all primitives along the ray.

\paragraph*{Volumetric representations of matter}
For heterogeneous media, discrete voxel-grid approaches are ubiquitously found in most applications due to their simplicity and flexibility. They can feature multi-level hierarchical structures~\cite{museth2013vdb, museth2021nanovdb}, which are helpful for faster traversal and filtering for level-of-detail applications, but suffer from poor scalability in terms of memory consumption when the modeled medium possesses finer details.
Sparser representations like collections of isotropic and simple volumes have been used in other fields, e.g., particle physics~\cite{Brown03}, where they can model neutron transport in graphite pebble-bed reactors. The advantage of having collections of these simple volumes is that, individually, they offer closed-form solutions for transmittance estimation and sampling, reducing variance when compared with most tracking methods normally used in voxel grids. However, searching for the boundaries of these volumes can be inefficient and slow in many situations~\cite{bitterli2018radiative}. In graphics, we can find hierarchical grid-based approaches~\cite{ProductionVolumeRendering2017} for subdividing complex, potentially sparse, volumes into smaller, more compact representations, though eventually they rely on smaller voxel-based primitives. Alternatively, implicit representations of participating media have been proposed, most notably in the context of radiance fields but also in compression of large volumes, including large multilayer perceptrons (MLP)~\cite{mildenhall2020,neuralvdb} or sparse hash grids of features combined with tiny MLPs~\cite{mueller2022instant}. These approaches result in large compression rates, at the cost of expensive queries.
Closer to our work, Knoll et al.~\shortcite{Knoll2021} proposed to model emissive and absorbing media using isotropic Gaussian mixture models, in the context of particle-based volumes~\cite{max1979atomlll}; however, they did not leverage closed-forms expressions for transmittance,  requiring stochastic integration via tracking, and cannot represent scattering media. In contrast, our work proposes closed-form solutions for transmittance and sampling, and further extends to support scattering media, anisotropic kernels, and efficient inverse rendering.

\paragraph*{Inverse volumetric rendering}
Early works on inverse rendering of heterogeneous media used inverse volumetric rendering for matching micron-scale cloth patches to measures~\cite{khungurn2015}, approximating multiscale volumetric representations~\cite{zhao2016downsample}, scattering compensation for 3D printing~\cite{elek17, sumin19, nindel2021gradient, Condor2023}, or inverse scattering~\cite{Gkioulekas2016AFF} by using hand-made derivatives of light transport. Zhang et al.~\shortcite{Zhang2019drt} proposed a differential radiative transfer framework suitable for inverse volumetric rendering. These works use voxel grids for representing media density, though our proposed representation would fit in these inverse pipelines.

\paragraph*{Radiance fields}
Neural Radiance Fields (NeRF)~\cite{mildenhall2020} introduced a continuous, implicit volumetric representation based on multi-layer perceptrons (MLPs). 
Subsequent works have leveraged multi-scale representations~\cite{mipnerf} and extended to unbounded scenes~\cite{barron2022mipnerf360, zhang2020nerf++, reiser2023merf}; however, training and evaluating the MLPs still requires significant computation.
In contrast, Instant-NGP~\cite{mueller2022instant} relies on a sparse hash grid of features and a small MLP, enabling faster scene reconstruction. Sparsification and hierarchization has been a popular avenue for optimizing radiance field fitting and rendering~\cite{yu2021plenoctrees, hedman2021snerg, bovzivc2022neu=ral, chen2022mobilenerf, Lombardi21}, achieving real-time rendering even on mobile devices, or helping on accelerating off-line rendering with complex assets~\cite{zhu2021neural,condor22}.
Most recently, 3D Gaussian Splatting~\cite{kerbl20233Dgaussians} removes the need for MLPs altogether, relying instead on an explicit Gaussian-based representation.
Combined with an efficient splatting technique for rasterization~\cite{ewasplattingzwicker04}, their approach is able to achieve fast rendering performance and high-quality reconstruction, allocating more primitives in areas where finer detail is needed. Our work takes Kerbl's approach, and extends it to general scattering media in the context of the radiative transfer framework.
\paragraph*{Gaussians in physically-based rendering}
Gaussian mixture models defined over the sphere have been extensively used in graphics for fitting BRDFs or environment maps~\cite{Xu13sigasia}, precomputed radiance transfer~\cite{green2006view}, or on-line path guiding~\cite{vorba2014line}. Yan et al.~\shortcite{yan2016position} fitted a 4D Gaussian mixture model for accelerating computation when rendering glinty surfaces. Jakob et al.~\shortcite{jakob2011progressive} proposed to fit a 3D anisotropic Gaussian mixture model to the distribution of radiance in media, showing accelerated rendering using photon-based techniques. Despite their work sharing some similarities to ours (closed-form integration along Gaussians), it focuses on fitting radiance in the context of radiance estimation, while our approach focuses on modeling matter (density) and can be used in arbitrary volumetric renderers. Combining both representations (kernel-based representations for both media and radiance) would be an interesting line of future work.

\section{Background}
\label{sec:background}
Under the geometric optics assumption, light transport in participating media is governed by the radiative transfer equation (RTE)~\cite{chandrasekhar1960radiative}. We compute the incident radiance $\sRad(\px,\wo)$ at point $\px$ and direction $\wo$ as
\begin{align}
    \label{eq:rte}
    (\wo\cdot\nabla)\sRad(\px,\wo) &= -\sAbs(\px) \sRad(\px,\wo) -\sScatt(\px) \sRad(\px,\wo) \\
    & + \sAbs(\px)\sRadE(\px,\wo) + \sScatt(\px) \sRadS(\px,\wo), \nonumber
\end{align}
where $\sRadS$ is the in-scattering term, defined as 
\begin{equation}
    \sRadS(\px,\wo) = \int_\Sphere \sPF(\px, \wi\rarrow\wo) \sRadI(\px,\wi)\, \diff\wi, 
\end{equation}
that models the light scattered in direction $\wo$ as a function of the integral of the light incoming from all directions on the unit sphere $\Sphere$.
The interaction of light and matter is characterized by the optical properties of the medium, described in terms of the differential probabilities of absorption and scattering  (or coefficients) $\sAbs(\px)$ and $\sScatt(\px)$ respectively, the extinction coefficient $\sExt(\px)=\sAbs(\px)+\sScatt(\px)$, the phase function $\sPF(\px, \wi\rarrow\wo)$, and the emission $\sRadE(\px,\wo)$. 
Assuming uncorrelated statistics of matter, the differential probability of extinction is defined as the product between the density $\sDensity(\px)$ of the particles forming the medium at point $\px$ [m$^{-3}$] and the particles' cross section $\sCross(\px)$ [m$^{2}$] as $\sExt(\px) = \sDensity(\px)\,\sCross(\px)$. 
For simplicity, we ignore the wavelength dependency in Equation~\eqref{eq:rte}, assume scalar elastic light-matter interactions, and omit the potential anisotropy of the medium~\cite{jakob2010radiative}; generalizing the following derivations to include these additional dimensions is trivial.
By integrating both sides of the differential RTE~\eqref{eq:rte} along the direction $\wo$ we obtain the volumetric rendering equation, that models radiance $\sRad(\px,\wo)$ as
\begin{align}
    \label{eq:vol_rendeq}
    \sRad(\px_0,\wo) & = \int_0^s \sTrans(\px_0,\px_t)\,\sExt(\px_t)\,\sRadO(\px_t, \wo)\,\diff t, \nonumber \\
    & + \sTrans(\px_0,\px_s) \sRad(\px_s,\wo)  \qquad\text{ with } \\
    \label{eq:vol_radout}
    \sRadO(\px, \wo) & = \left(1-\sAlbedo(\px)\right) \sRadE(\px,\wo) + \sAlbedo(\px) \sRadS(\px,\wo),
\end{align}
with $\px_t=\px_0 - t\wo$ and distance $t$, $\sAlbedo(\px)=\sScatt(\px)/\sExt(\px)$ as the single scattering albedo, and $\sTrans(\px_0,\px_t)$ is the transmittance defined as the fractional visibility due to absorption and out-scattering in the medium, and modeled following the Beer-Lambert law as
\begin{equation}
    \sTrans(\px_0,\px_t) = \exp\left(-\int_0^t \sExt(\px_{t'}) \diff t'\right).
    \label{eq:transmittance}
\end{equation}
Finally, $\sRad(\px_s,\wo)$ is the outgoing radiance at the medium boundary $\px_s$ at distance $s$ modeled by the rendering equation~\cite{kajiya86}.

\section{Volumetric Primitives}
\label{sec:method}
Inspired by recent works using Gaussian primitives for reconstructing radiance fields~\cite{kerbl20233Dgaussians}, and following a similar medium definition as Knoll et al. \shortcite{Knoll2021}, we model media using sets of primitives. Each primitive $\sPrim_i$  represents a statistical aggregate of matter with identical emission, cross section and phase function, and with density $\sDensity_i(\px)$ defined by a three-dimensional un-normalized kernel $\sKernel_i(\px)$. 
Given these primitives, we can model the distribution of matter, and therefore the extinction probability as a mixture of these volumetric primitives following
\begin{align}
    \sExt(\px) = \sum_{i=1}^N \sCross_i \, \sKernel_i(\px),
    \label{eq:extinction}
\end{align}
with $N$ the number of primitives affecting $\px$, and $\sCross_i$ their cross-section. For kernels with infinite support, $N$ is the total number of primitives $N_\text{total} = N$. In our case, we assume that all kernels have limited support, so typically $N<N_\text{total}$. We compute the single-scattering albedo and phase function at $\px$ analogously, with the key difference of requiring normalization (see Appendix~\ref{app:opt_params}).

\begin{figure}
\includegraphics[width=\columnwidth]{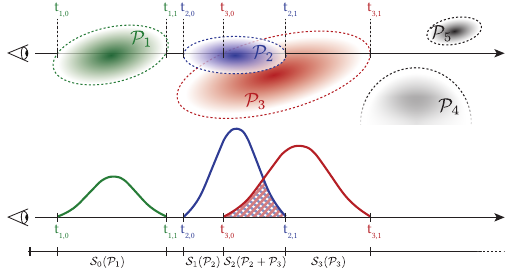}
\caption{\textbf{Top:} Flatland representation of a medium modeled using $N_\text{total} = 5$ primitives $\sPrim_i$ with $i\in [1,5]$. For the depicted ray from the camera only three primitives contribute directly to the ray. \textbf{Bottom:} The primitives are projected in 1D along the ray, and their density, emission and in-scattering are integrated over that line, which can be separated in disjoint segments defined by the boundaries of the primitives. The contribution of each segment $\sSegment_k$ is the integral over the primitives overlapping on that segment. }
\label{fig:integral}
\Description[<short description>]{<long description>}
\end{figure}

\subsection{The RTE with volumetric primitives}
Defining matter as a mixture of volumetric primitives with finite support allow us to reinterpret Equation~\eqref{eq:vol_rendeq} as the summation of the contribution of all the primitives. For that, as shown in Figure~\ref{fig:integral} (bottom), we split the ray in segments based on the primitives' entry and exit points; then, we integrate radiance along segments in a front-to-back pass, keeping track of the multiplicative transmittance which is computed analytically. This approach is similar to how production renderers deal with complex overlapping media~\cite[Ch.6]{ProductionVolumeRendering2017}; the key difference is that our approach approximates the media with primitives, which allows having closed-form transmittance and sampling expressions.

More formally, the boundaries of each primitive $t_{i,0}$ and $t_{i,1}$ subdivide the ray into ordered segments $\sSegment_k = [\stsegment_{k,0}, \stsegment_{k,1}]$, with $\stsegment_{k,0}=\stsegment_{k-1,1}$ for $k>0$, with $\stsegment_{0,0} = \px$. Each of these segments $\sSegment_k$ might overlap with zero, one, or multiple primitives denoted with the set $\{\sPrim_i | i\in \sSegmentSet_k\}$, where $\sSegmentSet_k$ is the per-segment set of indices, with $|\sSegmentSet_k|\leq N$ the number of primitives overlapping in segment $\sSegment_k$. 
With that definition of segments along the ray, we rewrite Equation~\eqref{eq:vol_rendeq} (omitting the boundary condition at $\px_s$ for simplicity, as
\begin{equation}
    \sRad(\px_0,\wo) = \sum_{k=1}^M \sTrans_{k-1}(\px_0,\px_t) \sRad_k(\px_{\stsegment_{k,0}},\wo), 
    \label{eq:segments_rte2_main}
\end{equation}
where $M$ is the number of segments along the ray, $\sRad_k(\px_{\stsegment_{k,0}},\wo)$ the outgoing radiance at the segment $\sSegment_k$ defined as
\begin{equation}
\sRad_k(\px_{\stsegment_{k,0}},\wo) = \int_{\stsegment_{k,0}}^{\stsegment_{k,1}} \sTrans(\px_{\stsegment_{k,0}}, \px_t) \, \sum_{i \in \sSegmentSet_k} \left[ \sCross_i \sKernel_i(\px_t) \, \sRadOI(\px_t,\wo) \right]\,\diff t,
\label{eq:segment_contribution}
\end{equation}
with $\sRadOI(\px,\wo)$ the outgoing radiance from primitive $\sPrim_i$, and $\sTrans_{k-1}(\px_0,\px_t)$ is the transmittance defined as a recursive operator
\begin{equation}
    \sTrans_k(\px_0,\px_t) = \sTrans_{k-1}(\px_0,\px_{\stsegment_{k-1,1}})\,\sTrans(\px_{\stsegment_{k,0}}, \px_t),
    \label{eq:rec_transmittance}
\end{equation}
with $\sTrans_{0}(\px_0,\px_t) = 1$ and  $\sTrans_M(\px_0,\px_t)=\sTrans(\px_0,\px_t)$. Trivially, transmittance in segment $k$ is computed as the product of transmittance of all primitives overlapping the segment 
\begin{equation}    
\sTrans(\px_{\stsegment_{k,0}}, \px_t)=\exp\left(-\sum_{i\in\sSegmentSet_k} \sOptDepth_i\left(\px_{\stsegment_{k,0}},\px_{\min(t,\stsegment_{k,1})}\right)\right), 
\end{equation}
with $\sOptDepth_i(\px_a, \px_b)$ the optical depth from primitive $\sPrim_i$ in the range $t\in[a,b]$ defined as
\begin{equation}
    \sOptDepth_i(\px_a,\px_b) = \sCross_i \, \int_{\max(a,\,t_{i,0})}^{\min(b,\,t_{i,1})} \sKernel_i(\px_{t'}) \, \diff t'.
    \label{eq:primi_optdepth}
\end{equation}
By choosing the appropriate primitive kernels, we show in Section~\ref{sec:kernels_gaussian} that this integral can be computed in closed form. A more detailed derivation can be found in the Supplemental material.
\subsection{Solving Equation~\ref{eq:segments_rte2_main} for distance sampling}
\label{sec:solving_prte}
While Equation~\eqref{eq:segments_rte2_main} is just a summation over segments that can be computed analytically (we show examples later in Section~\ref{sec:applications}), in the most general case $\sRadOI(\px_t,\wo)$ hides the inscattering integral, which requires numerical evaluation using Monte Carlo sampling. Computing a Monte Carlo estimate would quickly become impractical; thus we need to sample a single segment per ray, which requires essentially sampling the distance $t$ along the ray, ideally with probability distribution function (PDF) $\sPDF(t) = \sExt(\px_t) \sTrans(\px_0, \px_t)$.

While our primitive-based media allows the use of delta tracking (or any other Monte Carlo-based) estimators for distance sampling, we found that we can leverage the recursive formulation of transmittance~\eqref{eq:rec_transmittance} and pose the problem as an iterative search problem, where we uniformly sample the transmittance with a random variable $\xi\in(0,1)$, and search for the distance $t$ so that $\xi(t) = \sTrans(\px_0,\px_t)$. This approach is very similar to regular tracking~\cite{stutton99regular}, which we extend to handle overlapping kernels efficiently. In particular, we first search for a segment $\sSegment_k$ so that $(1-\xi)\in[\sTrans_{k-1},\sTrans_{k}]$. 
Then, we invert the transmittance inside the segment, solving for $t\in[\stsegment_{k,0},\stsegment_{k,1}] $ the following equation: 
\begin{equation}
    \log(1-\xi) = -\sum_{i \in \sSegmentSet_k} \sOptDepth_i\left(\px_{\stsegment_{k,0}},\px_{t}\right).
    \label{eq:trans_inversion_main}
\end{equation}
Depending on the type of kernel $\sKernel_i(\px)$ being used, and the amount of overlaps in segment $\sSegment_k$ this equation may have a closed-form analytical solution. However, for cases where Equation~\eqref{eq:trans_inversion_main} does not have a simple analytical solution (e.g. many overlapping kernels at the same segment), we rely on numerical root-finding. In particular, we use two different methods depending on the kernel: the Newton-Raphson and the bisection solvers. The former is our primary choice, since it is efficient and converges with good precision with a few iterations (1-3), given that our problem is well-conditioned. However, for certain long-tailed kernels (e.g., Gaussian) it occasionally suffers from numerical instability. In these cases, we rely in the slightly slower but more stable bisection method. Details for both methods can be found in the Supplemental.

\paragraph{Biased uniform distance sampling inside segments}
To maximize speed, in complex assets fitted with a high number of small kernels we can avoid inversion~\eqref{eq:trans_inversion_main} and simply sample inside each segment uniformly. While this means we sample with pdf $\sPDF(t) \neq \sExt(\px_t) \sTrans(\px_0, \px_t)$, in our implementation we assume that the probability distribution is close enough to it, allowing us to cancel out transmittance when computing the path throughput. This introduces a small amount of bias in exchange for a significant performance boost, which we analyze in Section~\ref{sec:forward_rendering}.
\section{Kernels}
In the previous section, we have derived a general framework for radiative transport in primitive-based media, defined using arbitrary kernel functions. Any kernel could be used within our framework as long as 1) they have limited support or their decay is such that they can be bounded or clipped efficiently and 2) closed-form solutions to their line integrals exist or can be numerically computed efficiently.

\begin{figure}
    \centering    
    \includegraphics[width=\columnwidth]{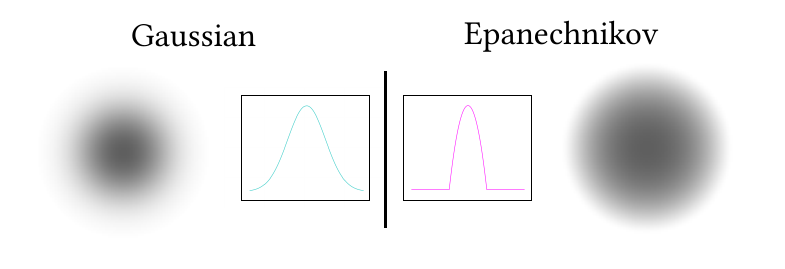}
    \caption{\revision{Visual comparison between the 1D (insets) and 2D Gaussian and Epanechnikov kernels. We plot $\pm{}3\sigma$ support to visualize the main difference between them: while the Gaussian kernel has infinite support, the Epanechnikov is finite, and decays sharply to 0. We further analyze differences between these kernels in Section~\ref{sec:kernelcomp}}.}
    \label{fig:kernel_comp_plots}
    \Description[<short description>]{<long description>}
\end{figure}

Here we implement our framework using two different kernels: the Gaussian and the Epanechnikov kernel. These have been successfully used in the signal processing, density estimation, rendering, and inverse graphics literature~\cite{liu2021epan, jakob2011progressive, kerbl20233Dgaussians}. Figure~\ref{fig:kernel_comp_plots} shows their shape; we compare their performance in different applications in Section~\ref{sec:kernelcomp}.

\subsection{Gaussian Kernel}
\label{sec:kernels_gaussian}
The normalized Gaussian kernel $g_i(\px)$ for primitive $\sPrim_i$ is defined as 
\begin{equation}
\label{eq:3dgaussian}
    g_i(\px) = \frac{1}{(2\pi)^\frac{3}{2}(|\Sigma_i|)^\frac{1}{2}} e^{-\frac{1}{2}(\px-\bm{c}_i)^T\Sigma_i^{-1}(\px-\bm{c}_i)},
\end{equation}
where $\Sigma_i$ is the $3\times3$ covariance matrix and \revisionB{$\bm{c}_i$} its mean. As opposed to other convex kernels, Gaussians have unlimited support, which would potentially require evaluating all primitives in the scene for every ray. In practice, similar to Kerlb et al.~\shortcite{kerbl20233Dgaussians}, we bound the kernel's support to three times the standard deviation, covering 99.73\% of the Gaussian's density. 

\paragraph{Transmittance Evaluation}
We compute the closed-form transmittance of a Gaussian primitive $\sPrim_i$ from point $\px_{t_0}$ to point $\px_{t_1}$ along the ray $\ray(t)=\px_0 + \dir\,t$, by integrating the optical depth $\sOptDepth_i(\px_{t_0},\px_{t_1})$. These parameters are them defined in the Gaussian's local coordinate system, also known as whitened space (which we obtain by linearly transforming the ray into the canonical space of the Gaussian ellipsoidal shell, where the ellipsoid is a sphere of radius 1). Due to the symmetry of the Gaussian kernel around its mean, we can simplify the definite integral by shifting the origin of the ray to ensure a symmetric integration domain, integrating optical depth $\sOptDepth_i(\px_{-t_s},\px_{t_s})$ with $t_s = \frac{t_0 + t_1}{2}$. This reduces the number of error functions required from 2 to 1. Plugging in Equation~\eqref{eq:3dgaussian} into Equation~\eqref{eq:primi_optdepth} we get
\begin{align}
\label{eq:opt_depth_gaussian}
    \sOptDepth_i(\px_{-t_s},\px_{t_s}) & = \sCross_i \int_{-t_s}^{t_s} g_i(\px_t) \,\diff{t} = G
    e^{ -\frac{1}{2} \left(a - b^2 \right)} 
    \left[\operatorname{erf} \left( t_s\sqrt{\frac{1}{2}} \right) \right]
\end{align}

with

\begin{equation}
    \begin{aligned}
        a = \px_{0} \times \px_{0} , \quad
        b = \px_{0} \times \dir , \quad
        G = \frac{\sCross_i}{2\pi\sqrt{v^T\Sigma_i^{-1}v|\Sigma_i|}}
    \end{aligned}
\end{equation}

with $\erf(\cdot)$ being the error function. Assuming exponential media, we can compute the transmittance as $T(\px_{t_0}, \px_{t_1})=\exp(-\sOptDepth_i(\px_{t_0},\px_{t_1}))$.

\paragraph{Distance Sampling}
For sampling the free-flight distance $t$ in a single Gaussian primitive $\sPrim_i$ with PDF $\sPDF(t) = \alpha_i G_i(\px_t) \sTrans(\px_0, \px_t)$, we need to invert its optical depth~\eqref{eq:opt_depth_gaussian}, which by setting $t_0 = 0$ and using a random value $\xi\in(0,1)$, and assuming an accumulated transmittance up to $t_0$ of $\beta$ has closed form following 
\begin{align}
\label{eq:inv_cdf}
    t(\xi) = \sqrt{2} \operatorname{erf}^{-1} \left( \operatorname{erf}\left( \frac{b}{\sqrt{2}} \right) - \frac{2\operatorname{log}(\xi/\beta)}{Ge^{-\frac{1}{2}(a - b^2)}} \right) - b \\ 
\end{align}
For segments where more than one 3D Gaussian is contributing, we revert to using the bisection solver presented in Section~\ref{sec:method}. For single kernels, we normally use Equation~\ref{eq:inv_cdf}. In practice, the piece-wise approximation of the $\erf^{-1}$ function commonly implemented in GPU math libraries~\cite{cuda} produces large errors for certain input values, and for these we resort to the bisection solver as well, at the cost of some performance. 

\subsection{Epanechnikov Kernel}
In our primitive-based rendering framework, Epanechnikov kernels have a significant advantage compared to Gaussian kernels due to their limited support. This allows more compact primitive shells, which accelerate rendering by reducing the number of overlaps.
\revision{The 3D Epanechnikov kernel $\mathcal{E}_i(x)$ for primitive $\sPrim_i$} is defined as

\begin{equation}
\label{eq:epa_kernel}
 \mathcal{E}_i(\px) = \begin{cases}
 \frac{15}{8\pi(7^3\left | \Sigma_i  \right |)^\frac{1}{2}}[1 - \frac{1}{7} d(\px)] & \text{ if } d(\px) \leq 1 \\
 0 & \text{ otherwise }, 
\end{cases}
\end{equation}
where, analogously to the Gaussian kernel, $\Sigma_i$ is the $3\times3$ covariance matrix, $\bm{c}_i$ the mean, and $d(\px) = (\px-\bm{c}_i)^T\Sigma_i^{-1}(\px-\bm{c}_i)$.

\paragraph{Transmittance Evaluation} We derive the closed-form transmittance of an Epanechnikov-based primitive between points $\px_{t_0}$ and $\px_{t_1}$ by integrating the optical depth between them as

\begin{align}
\label{eq:opt_depth_ep}
    \sOptDepth_i(\px_{t_0},\px_{t_1}) & = \sCross_i \int_{t_0}^{t_1} \mathcal{E}_i(\px_t) \diff{t} \\
    & = \sigma_i{}\mathcal{K}_{\text{norm}}(\mathcal{K}_3{}(t_0^3 - t_1^3)+\mathcal{K}_2{}(t_0^2 - t_1^2) +\mathcal{K}_1{}(t_0 - t_1)), \nonumber
\end{align}
where $\mathcal{K}_1$, $\mathcal{K}_2$, $\mathcal{K}_3$ and $\mathcal{K}_{\text{norm}}$ are constants that depend on the ray position and direction, and the eigenvalues of the kernel's covariance matrix $\Sigma_i$. Explicit expressions for these constants can be found in Appendix~\ref{app:kernel_transmittance}. Analogously to any other kernel, and assuming exponential media, we can finally compute the transmittance as $T(\px_{t_0}, \px_{t_1})=\exp(-\sOptDepth_i(\px_{t_0},\px_{t_1}))$.

\paragraph{Distance Sampling}
While there is an analytic solution for inverting Equation~\eqref{eq:trans_inversion_main} for one Epanechnikov kernel, it is not practical, given its complexity. Instead, we directly use the Newton-Raphson solver for segments with both one or multiple overlapping kernels.
\section{Implementation Details}
\label{sec:implementation_details_integrator}
We implement our method in Mitsuba 3~\cite{jakob2022mitsuba3}, which we extended to support volumes of kernel primitives with different shell geometries, where we use ray tracing for querying primitives along the ray. We implement two different integrators, depending on the target application: 1) a \emph{volumetric-primitives path tracer} (\emph{VPPT}) supporting scattering media, and 2) a simplified \emph{volumetric-primitives radiance field} (\emph{VPRF}) integrator that computes the radiance field along a primary ray.

In both cases, we compute transmittance and emission by following the segment-based formulation described in Section~\ref{sec:solving_prte}, by iterating over the segments in an ordered fashion: This fits very well with our ray-tracing-based querying of the primitives, since we collect new segments by simply casting new rays from the previous intersection. Listing~\ref{code:primitive-tracing} shows a skeleton pseudocode for our two integrators.
\subsection{Integrators}
\paragraph{Volumetric-primitives path tracer (VPPT)} This integrator is mostly an off-the-shelf volumetric path tracer supporting next-event estimation, with the key modifications of the transmittance evaluation and sampling, and the medium interaction routines. It solves Equation~\eqref{eq:segments_rte2_main} using Monte Carlo estimation over multiple paths where, assuming non-emissive media, the throughput of each sample path is recursively computed as
\begin{equation}
L(\px_0) = \beta(\px_0,\px_i) \sExt(\px_i) \sAlbedo(\px_i) \sRadS(\px_i, \omega_i),
\end{equation}
with $\omega_i=\frac{\px_i-\px_{i-1}}{|\px_i-\px_{i-1}|}$, and $\beta(\px_0\rightarrow\px_i)$ the throughput of the eye subpath until $\px_i$ defined as
\begin{equation}
    \beta(\px_0,\px_i) = \frac{\beta(\px_0,\px_{i-1}) \sTrans(\px_{i-1}, \px_i) \sExt(\px_{i-1}) \sAlbedo(\px_i) \sPF(\px_{i-1}, \omega_{i-1}\rarrow\omega_i)}{p(\px_i)},
    \label{eq:vppt}
\end{equation}
where $p(\px_i)$ is the probability of generating $\px_i$ from the previous scattering vertex $\px_{i-1}$. In our implementation, we obtain $\px_i$ using the standard practice of first sampling the phase function at $\px_{i-1}$, and then sampling distance with probability proportional to $\sTrans(\px_{i-1}, \px_i)$. 

\paragraph{Volumetric-primitives radiance field (VPRF)} This implements in our volumetric primitives framework the radiance field image formation model, similar to e.g.~\cite{mildenhall2020,kerbl20233Dgaussians}. It significantly simplifies the image formation model from VPPT, by only considering the primary rays from the camera, accumulating radiance along the ray by querying a spherical harmonics-based emission in the primitives. In addition, we disable overlapping logic in order to favour speed. 
In practice, this means that we approximate Equation~\eqref{eq:segments_rte2_main} using an ad-hoc approximate of the emission at each segment following
\begin{equation}
    \sRad_k(\px_{\stsegment_{k,0}},\wo) \approx \left(1-\sTrans_k(\px_{\stsegment_{k,0}}, \px_{\stsegment_{k,1}})\right) \sum_{i \in \sSegmentSet_k} \frac{\sOptDepth_i\left(\px_{\stsegment_{k,0}}, \px_{\stsegment_{k,1}}\right)}{\sum_{j \in \sSegmentSet_k} \sOptDepth_j\left(\px_{\stsegment_{k,0}}, \px_{\stsegment_{k,1}}\right)} \sRedRadE_i(\wo).
    \label{eq:rad_segment_rf}
\end{equation}
\subsection{Ray tracing volumetric primitives}
As described above, we limit the support of our kernels by bounding their bandwidth with an ellipsoid shell. We query the primitives along a ray leveraging ray-tracing, which we accelerate through the construction of a BVH using Optix~\cite{optix} for efficient ray traversal. We implemented a ray-ellipsoid intersection as a custom intersection routine in Optix. However, we found that this produces suboptimal BVHs given the potentially large support of the primitives. Moreover, a custom intersection kernel is significantly less performant than a hardware-accelerated ray-triangle intersection test. Thus, instead of using the ray-ellipsoid intersection, we can triangulate the ellipsoids, and ray-trace on triangles. We analyzed a variety of triangulation approaches (see Figure~\ref{fig:shell_ablation}), and found a significant boost on performance despite the higher number of primitives in the scene, finding highly-tesselated icospheres to perform best in terms of execution times. One could consider using mesh instancing to avoid duplicating the shell's geometry. However, this approach reintroduces the problem of using axis-aligned bounding boxes for the individual instances in the instances BVH.

\paragraph{Stack array allocation}
A limitation of our \textit{primitive tracing} algorithms is their need to stack information while iterating over the primitives along the ray. This is particularly necessary for calculating the list of primitive exit points during segment iteration, as well as maintaining a record of the active primitives for the current segment. Unfortunately, most JIT frameworks, such as Dr.Jit or PyTorch, do not offer an API with a low enough level to perform pointer arithmetic. For efficiency, it is crucial to execute this performance-critical algorithm in registers, which necessitates the ability to allocate, access, and write into arrays of variables on the stack rather than on the heap. Fortunately, the NVidia PTX assembly language offers local-state space private memory for each thread to store its own data, which can be utilized for this purpose. We modified the Dr.Jit~\cite{jakob2020drjit} core to include routines in the JIT compiler that generate such intrinsics.

\begin{code}
\begin{minted}[linenos,fontsize=\footnotesize, escapeinside=||, mathescape=true]{python}
def primitive_tracing(ray, max_depth, P = []):
  depth = 0
  |$t_0$| = 0.0 # Current segment start time
  while depth < max_depth:
    V = [(p, p.|$t_1$|) for p in P] # List primitive exit points
    p = ray_intersect(ray)     # Find next primitive
    V.append((p, p.|$t_0$|))        # Add primitive entry point

    # Process all exit points up to this intersection
    while not V.is_empty():
      # Find next vertex to process
      v = min(V, key=lambda x: x.|$t$|)

      # Process segment according to application
      process_segment(ray, G, |$t_0$|, v.|$t$|)

      |$t_0$| = v.|$t$| # Move on to next segment
      if p == v.p: break # Check if we have reached p
      P.remove(v.p) # Exiting v.p
      V.remove(v)

    P.append(p) # Ray is now entering primitive p
    ray = ray.move_to(p.|$t_0$|) # Update ray position
    depth += 1
\end{minted}
\caption{
    \label{code:primitive-tracing}
    Pseudo-code implementation of our primitive tracing algorithm. In \textit{VPPT}, \texttt{process\_segment()} samples a medium interaction in that segment and recursively calls \texttt{primitive\_tracing()} on sampled position. In \textit{VPRF}, \texttt{process\_segment()} computes the radiance field contribution of that segment following Equation~\eqref{eq:rad_segment_rf}.
}
\end{code}

\subsection{Differentiating volumetric primitive integrators}
We develop a backward derivatives propagation routine for our integrators, so that they can be used to solve inverse problems. Drawing parallels to the work by Vicini et al.~\shortcite{Vicini2021PathReplay}, these routines utilize the same sequence of random numbers to generate identical light paths in both the primal and backward rendering phases, and propagate the gradients to and from the scene parameters along these paths. Additionally, they accept the total radiance along these paths as input, which is then used to retrieve the various quantities required for reverse-mode differentiation.

A unique aspect of our adjoint implementation is the inclusion of loops to iterate over primitives along a ray and influence the different segments. Unfortunately, automatic differentiation systems like Dr.Jit~\cite{jakob2020drjit} do not support automatic derivative propagation through loops. Although it is possible to modify Dr.Jit to automatically generate the appropriate transformation, it would not yield an efficient outcome, as each loop iteration would require storing copies of all loop variables to facilitate a reversal under general conditions. Consequently, it is imperative to provide an adjoint formulation of the derivative propagation within those loops. 

Appendix~\ref{sec:supp_adjoint} includes the adjoint form for our two integrators. The full derivations and implementation details can be found in the Supplemental document, as well as comparisons against finite differences. It is important to highlight that the backward propagation process in our implementation has the capability to propagate gradients to all primitive parameters simultaneously. This is a significant advantage over finite differences, which typically handles parameters individually.
\section{Applications}
\label{sec:applications}
We can use our volumetric primitives and rendering algorithms in a wide range of volume rendering applications, from traditional forward rendering of scattering media, to physically-based (PB) inverse rendering, and radiance field optimization and rendering.

\subsection{Forward Rendering}
\label{sec:forward_rendering}
In our forward rendering experiments we focus on scenes where only density varies, and we fix the single-scattering albedo and phase function. We set the primitive kernel to a Gaussian kernel, which is more suitable for representing smooth volumetric data such as clouds or smoke.
We use three different volumetric assets, obtained from voxel grid representations. We transform the grid-based representation to a Gaussian mixture model (GMMs) by optimizing the GMM using the inverse tomography pipeline we describe later in Section~\ref{sec:inverse}, where the GMM is optionally initialized using expectation-maximization~\cite{EM}. The cost of fitting depends on the number of primitives, ranging from 15 minutes of the \emph{Smoke} asset (Figure~\ref{fig:absorptive}), to several hours in the most complex \emph{Cloud} asset (Figure~\ref{fig:teaser}). Note however that in our implementation this process is not optimized and can be significantly accelerated. Further details can be found in the Supplemental document.

\subsubsection{Experiment setup}
As baseline, we compare our method against a grid-based representation of the volume rendered using a weighted delta tracking free-flight sampler~\cite{woodcock, raab2006unbiased} with transmittance computed using residual ratio tracking~\cite{novak14}. For both free-flight sampling and transmittance estimation, we use a local majorant precomputed in a coarser grid where each supervoxel stores local statistics (local majorant, minorant, and average density) over a $4^3$ voxels neighborhood. This supervoxel-grid is traversed during sampling using DDA~\cite{szirmay2011free}. 
Having tighter (local) majorants substantially improves free-flight sampling in complex heterogeneous media, reducing the probability of null-scattering events. We choose \emph{residual ratio tracking} as the transmittance estimator for next-event estimation in all the references, using the average local density as control coefficient; cost per sample for our DDA traversal-based \emph{residual ratio tracking} is substantially smaller than e.g. \emph{power-series} estimators~\cite{georgiev19}, which is typically favored for high albedo volumes, where long paths and higher sample counts are expected.
Unless otherwise stated, the reference voxel grid is computed by voxelizing our mixture model into a high-resolution grid, so that the distributions of density between the reference and our method are as close as possible, allowing per-pixel error comparisons in terms of computing gain.

As for our method, we use our \emph{VPPT} integrator with next-event estimation (NEE). We trade off quality for performance in our analytic transmittance estimates (Figure~\ref{fig:absorptive}) with Russian Roulette termination during NEE, and simplifying sampling by uniformly sampling a position along the critical segment instead of inverting the CDF over all intersecting Gaussians.

\begin{figure}
    \centering
    \includegraphics[width=\columnwidth]{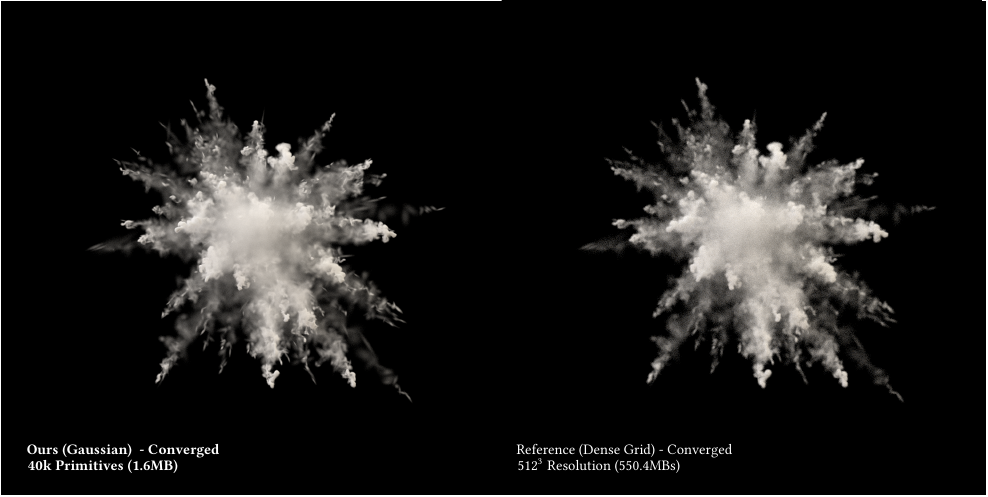}
    \caption{\emph{Dust explosion} asset, fitted to 40k Gaussians and discretized into a $512^3$ resolution grid, both rendered until convergence. We achieve high quality fits and renders on tiny memory budgets.}
    \label{fig:converged_dust}
    \Description[<short description>]{<long description>}
\end{figure}

\subsubsection{Results}
\revision{Figures~\ref{fig:teaser} and ~\ref{fig:converged_dust} demonstrate our method rendering complex assets with high constant single-scattering albedo ($\sAlbedo=0.99$). The main source of variance in these scenarios is the free-flight sampling.
Our method compares positively against state-of-the-art techniques for grid-based volume rendering, both in terms of performance and memory footprint. Given that the main limitation in production volume renderers has traditionally been memory bandwidth, our approach could deliver substantial speedups; while also being appealing in more memory-constrained environments.}

\begin{figure}
    \centering
    \includegraphics[width=\columnwidth]{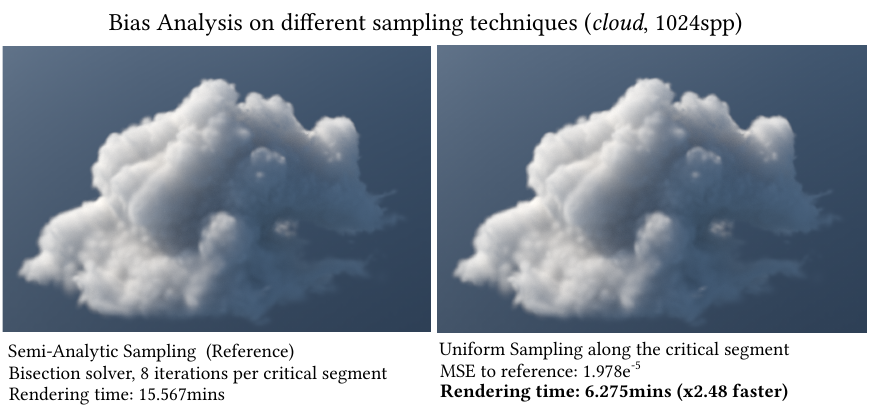}
    \caption{\emph{Cloud} asset, rendered with 1024 samples per pixel with both our unbiased semi-analytic (left) and our biased uniform sampling (right) distance sampling techniques inside each segment. Given the accuracy of our transmittance estimates and the small size of segments in complex assets such as \emph{cloud}, the bias introduced is minimal, while obtaining significant speed-ups. Rendering times reported on a NVIDIA RTX 3080.} 
    \label{fig:bias_comparison}
    \Description[<short description>]{<long description>}
\end{figure}

\revision{Figure~\ref{fig:bias_comparison} analyzes the bias introduced by using the biased uniform distance sampling inside the segment}, instead of numerically inverting the CDF. At the cost of a small amount of bias, uniformly sampling the distance inside segments gives a very substantial speed-up compared to the exact semi-analytical CDF inversion.

\begin{figure*}
    \centering
    \includegraphics[width=\textwidth]{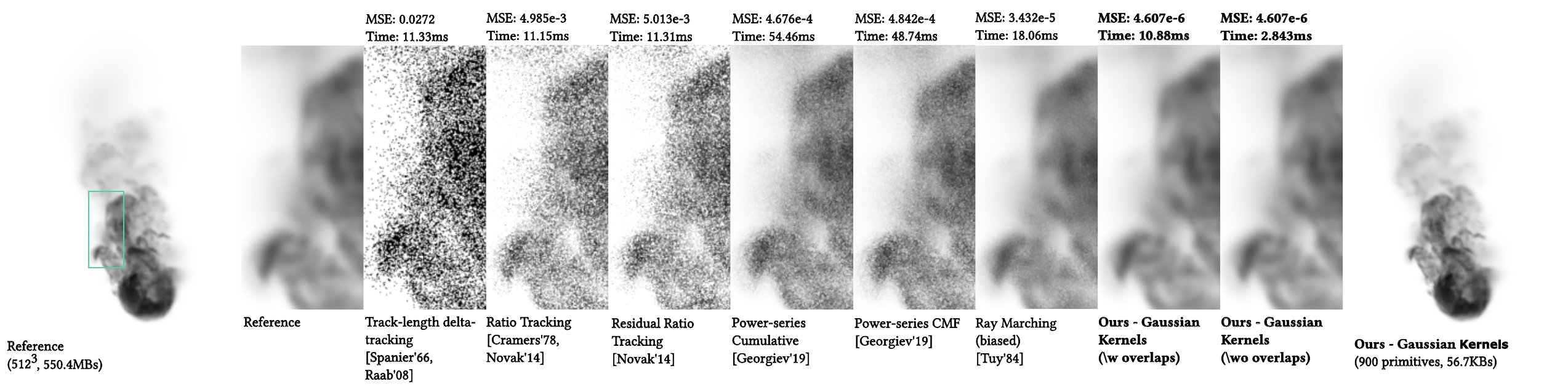}
    \caption{\emph{Smoke} asset, fitted to 900 Gaussians (37KB) and discretized into a $512^3$ resolution grid (550.4MBs) to be used by previous work. The supervoxels grid has a resolution of $64^3$ voxels. In this integrator, only primary rays are traced. We show a single sample per pixel (spp) in all the renders above. Our analytic transmittance estimation produces no variance from a single spp (reported error is primarily the discretization difference between the GMM and voxel grid distributions), while our rendering solution still delivers lower cost per sample than competing Monte Carlo-based estimators, at a fraction of the memory footprint. We can further reduce cost per sample to 2.84ms by disabling segment-by-segment integration, and simply integrating every intersected Gaussian along the primary ray paths as a whole, producing the same image. We use the former during distance sampling, and the latter during next-event estimation, maximizing efficiency while staying unbiased. Rendering times reported on a NVIDIA RTX 3080.}
    \label{fig:absorptive}
    \Description[<short description>]{<long description>}
\end{figure*}

\paragraph{Comparison against transmittance estimators}
\revision{In Figure~\ref{fig:absorptive} we compare our analytical transmittance estimation against state-of-the-art tracking techniques on an absorbing medium lighted by a constant white environment, where the sole contributor to variance is transmittance. We use the \emph{Smoke} asset, approximated with 900 Gaussians. We compare against a wide range of transmittance estimators: \emph{track-length delta tracking} \cite{spanier1966two, spanier2008monte, raab2006unbiased}, \emph{ratio tracking}\cite{cramer1978application, novak14}, \emph{residual ratio tracking} \cite{novak14}, \emph{ray marching} \cite{tuy1984direct, kenperlin89} and both \emph{power series-cumulative} and \emph{power series-CMF} \cite{georgiev19}. Note that all of these estimators are unbiased except \emph{ray marching}. 
For \emph{residual ratio tracking}, we use the average local density as the control coefficient, and both this estimator, \emph{ratio tracking} and \emph{track-length delta tracking} leverage local statistics to reduce cost per sample and improve sampling efficiency. The rest of the estimators use global statistics instead. Our analytic transmittance integrator produces estimates with zero variance even on a single sample per pixel. Our cost per sample is also the smallest of all tested approaches, which further showcases the efficiency of our approach.}

\begin{figure*}
    \centering
    \includegraphics[width=\textwidth]{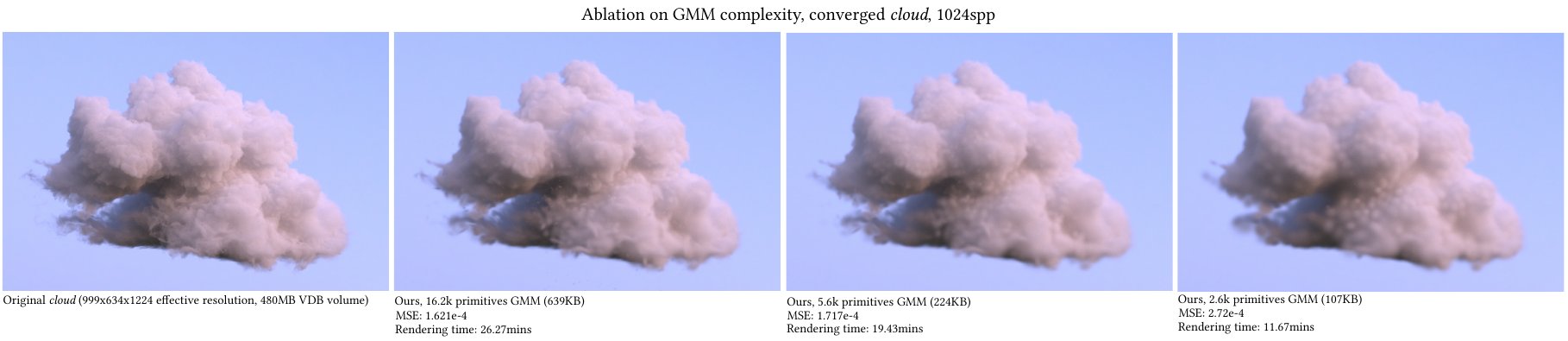}
    \caption{\emph{Cloud} asset, rendered at 1024spp with varying amounts of Gaussian primitives, vs the original VDB volume. We achieve high compression rates (a $\times5251$ compression rate with respect to the dense voxel grid, compared to the $\times6.823$ rate obtained by OpenVDB) and can easily trade off quality for memory and performance. Reported timings on a NVIDIA RTX 3080, equal settings for our method. Further performance gains can be achieved by tuning down individual parameters depending on the asset complexity (e.g., reducing the maximum number of overlapping kernels to be considered per segment).}
    \label{fig:gaussian_converged_comp}
    \Description[<short description>]{<long description>}
\end{figure*}

\paragraph{Analysis on quality-performance trade-off}
The performance of our method is dependent on the number of primitives used to model the density distribution, and most importantly, on the amount of overlap between them. However, aggregation is straightforward when using Gaussian distributions; in Figure~\ref{fig:gaussian_converged_comp}, we explore the scalability of our approach: at the cost of some visual fidelity, we can drastically increase compactness and rendering speed. This also makes our representation very suitable for level-of-detail (LoD) applications.

Note that our approach provides a significant compression rate even when compared with state-of-the-art adaptive grid approaches (OpenVDB), with a small penalty on quality (sharpness). In terms of performance, adaptive grid approaches require tracking algorithms as discussed earlier and thus would suffer from similar increased variance compared with our method. It is expected though that NanoVDB would provide some speed-up compared with dense grids due to reduced GPU memory bandwidth, with additional penalties on lookups due to tree traversal.

\begin{figure}
    \centering
    \includegraphics[width=\columnwidth]{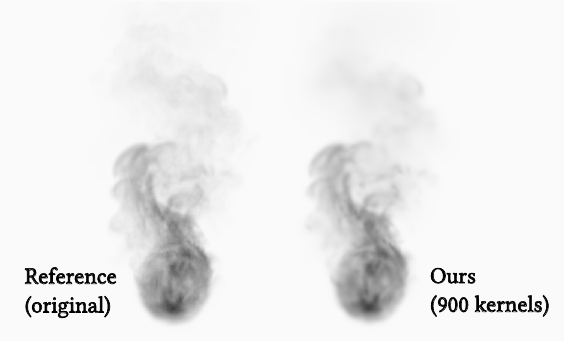}
    \caption{Original \emph{Smoke} voxel grid (left), and our fit with a 900 Gaussians computed using our inverse tomography pipeline (right), starting from a random set of primitives. We achieve a PSNR of 54.028dB and a compression rate over the original volume of $\times291$. We use this optimized asset in Figure~\ref{fig:absorptive}.}
    \label{fig:absorptive_opt_comp}
    \Description[<short description>]{<long description>}
\end{figure}

\subsection{Physically-Based Inverse Rendering}
\label{sec:inverse}

Our primitive-based formulation of media is very suitable for inverse problems where the optical properties of heterogeneous media are reconstructed from observations, given its compactness in both representation and gradients. For that, we use a differentiable version of our \emph{VPPT} integrator, with the adjoints presented in Section~\ref{sec:implementation_details_integrator}. In this section we demonstrate a proof-of-concept inverse pipeline, under different problems setups, including both the inversion of absorbing and scattering media. Note that it is outside the scope of this paper to provide a robust, fully-fledged optimization pipeline, which would require further investigation. 

\paragraph{Parameterization}
We parameterize our 3D Gaussians similarly to 3D Gaussian splatting~\cite{kerbl20233Dgaussians}, distilling them into mean, scale and rotation (using Euler angles instead of quaternions), which we optimize and then later use to compute the covariance matrix $\Sigma_i$ for the primitive $\sPrim_i$ as
\begin{equation}
    \Sigma_i = \mathcal{R}_i\mathcal{S}_i\mathcal{S}_i^T\mathcal{R}_i^T
\end{equation}
with $\mathcal{R}_i$ and $\mathcal{S}_i$ the primitive rotation and scale matrices, respectively.

\paragraph{Optimization Procedure}
We begin our set of primitives with a single entity and gradually incorporate more primitives by randomly sampling points on the bounding sphere of all current primitives. The parameters of the primitive are optimized using an adapted version of the Adam optimizer~\cite{kingma2017adam} that we call \textit{Bounded Adam}, which establishes hard limits on the boundary values of some parameters to avoid 1) intersection artifacts on extremely small primitives; 2) generating primitives outside of the determined scene bounding box, and 3) avoiding large primitives, which due to our reliance on axis-aligned bounding boxes for the BVH would increase our rendering times.

\paragraph{Bounded Adam} 
The key idea of Bounded Adam is to move the optimized value by half of the distance towards the bound if a gradient step reaches one of the bounds. This ensures that the optimization process can approach the bound as closely as possible without ever stepping over it. In our experience, this works better than simply clamping the values, as this process will also reset the optimizer state for this parameter. 

\paragraph{Loss}
We employ a $\lambda$-weighted combination of mean-absolute error (L1) and D-SSIM \cite{ssim}, to which we add two regularization terms to control anisotropy and density of small primitives. This prevents individual primitives from overfitting specific details in the training images. It should be noted that these regularization terms are only active above a certain threshold. In summary, the loss we use for reconstruction reads
\begin{equation}
    \mathcal{L} = \lambda \mathcal{L}_1 + (1-\lambda) \mathcal{L}_\text{SSIM} + \mathcal{L}_\text{ani} + \mathcal{L}_\text{d} \,,
\end{equation}
with 
\begin{equation}
    \mathcal{L}_\text{ani} = w_\text{ani} \, \sum_{i=1}^{N} \frac{\min\left(\text{Eig}(\mathcal{S}_i)\right)}{\max\left(\text{Eig}(\mathcal{S}_i)\right)} \qquad \text{and} \qquad \mathcal{L}_\text{d} = w_\text{d} \, \sum_{i=1}^{N} \frac{\sCross_i}{|\mathcal{S}_i|}, \nonumber
\end{equation}
where $w_\text{ani}$ and $w_\text{d}$ are the weight parameters controlling the strength of the two regularizers, and $\text{Eig}(\cdot)$ being the eigenvalue operator.

\subsubsection{Results}
\paragraph{Inverse tomography}
We first demonstrate our pipeline on reconstructing heterogeneous density on purely absorbing media, illuminated by a constant white environment with a Gaussian kernel. Figure~\ref{fig:absorptive_opt_comp} shows results on the \emph{Smoke} dataset, where we obtain a PSNR of 54.028dB with only 900 Gaussians, starting from a random set of Gaussians, from a set of 16 images around the smoke plume. Note that this pipeline has been used to generate all the assets in Section~\ref{sec:forward_rendering}.

\begin{figure}
    \centering
    \includegraphics[width=\columnwidth]{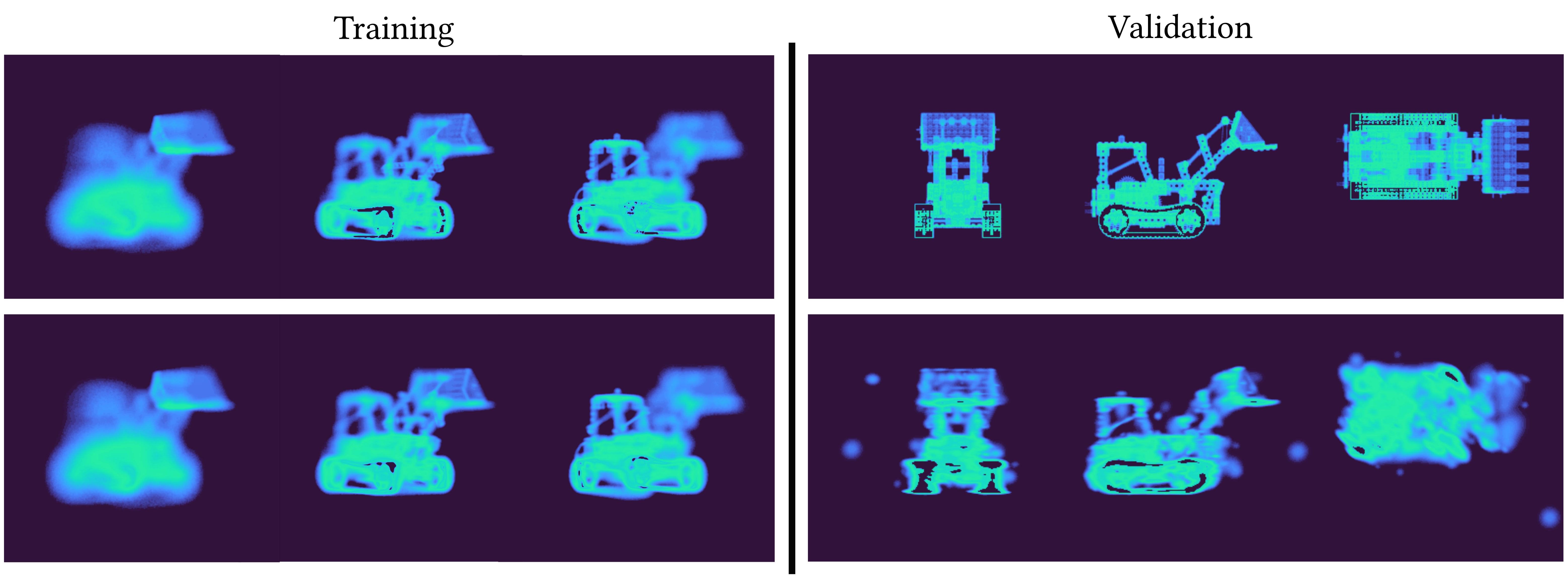}
    \caption{Inverse tomography from a focal stack. Left: Selected frames from a focal stack composed of 32 images, rendered with the same camera pose and linearly interpolating focus distance between far and near planes (top); fitting results after 1500 iterations (bottom, 907 primitives, PSNR: 39.3186). Right: Reference evaluation views (top) and novel rendered views (bottom) from our inverse telecentric tomographic reconstruction. We can see that even when trained on a single camera pose, it can reconstruct 3D structure.}
    \label{fig:focus_stack_inverse_tomography}
    \Description[<short description>]{<long description>}
\end{figure}

Figure~\ref{fig:focus_stack_inverse_tomography} further demonstrates our pipeline with a more challenging camera model, mimicking tomographic microscopy. We create a focal stack using a telecentric (orthographic thin-lens) camera from a single point of view on an absorbing \emph{Bulldozer} dataset. We reconstruct it with 907 Gaussians from that focal stack of 32 images using our pipeline. Even when employing a single point of view and using limited numbers of primitives, we  obtain good 3D reconstructions from the focus stack (PSNR: 39.3186), even from unseen points of view (right).

\begin{figure}
    \centering
    \includegraphics[width=\columnwidth]{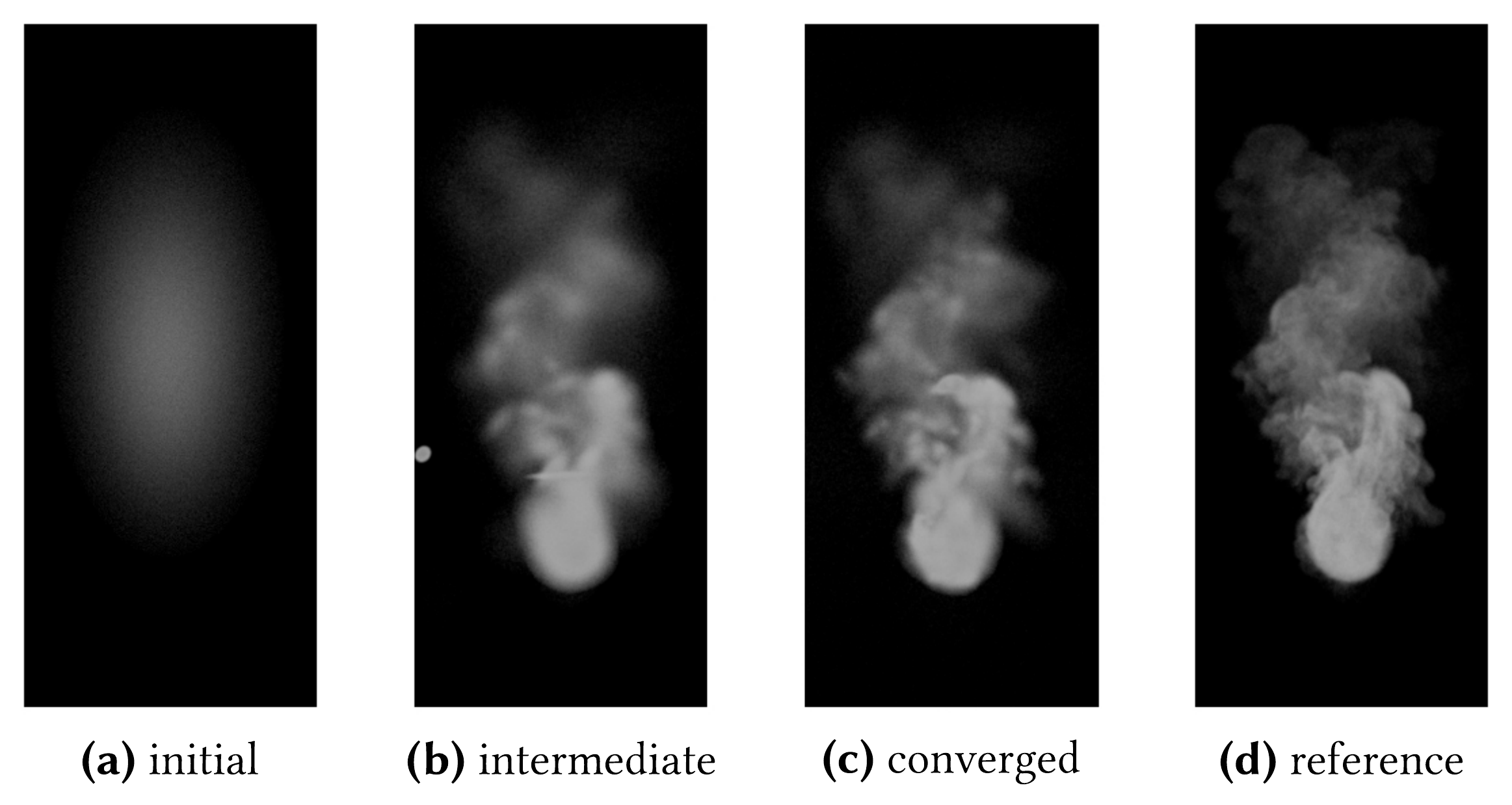}
    \caption{Inverse scattering results on the \emph{Smoke} asset. Starting from a single primitive (a), we iteratively optimize its parameters and spawn child primitives surrounding it, progressively adding more detail (b). Our final mixture model (c) achieves a PSNR of 36.64614 dB on average using only 551 Gaussians. The scattering medium is illuminated by a white enviroment, removed for visualization. }
    \label{fig:smoke_plume}
    \Description[<short description>]{<long description>}
\end{figure}

\paragraph{Inverse scattering}
\label{sec:inverse_scattering}
Figure~\ref{fig:smoke_plume} shows the result of a more complex experiment, where we reconstruct a heterogeneous \emph{scattering medium}. This is significantly more challenging than tomography, given the much higher dimensionality of the problem (i.e. long paths, high sampling variance due to multiple scattering). 
We reconstruct a highly-scattering version of the \textit{Smoke} plume with an isotropic phase function, from a set of 32 images (resolution of $256^2$ pixels) around the asset. As opposed to the inverse tomography in Figure~\ref{fig:absorptive_opt_comp}, we start from a single Gaussian and progressively spawn child Gaussians surrounding it in order to add detail in a progressive and controlled manner, and run our optimization for 1200 iterations. 
We converged to a 551-Gaussians fit, with a PSNR 36.64 dB on average. 

\subsection{Radiance Fields}
\label{sec:radiancefields}
As a third application, we present a complete inverse radiance field optimization and rendering pipeline, that reconstructs view-dependent appearance of both synthetic and real-world scenes from captured images. The purpose of this experiment is to further showcase the flexibility of our approach and demonstrate how it can enhance the capabilities of competing kernel-based radiance field solutions such as 3DGS.

For this application, we use our simpler \emph{volumetric primitive radiance field (VPRF)} integrator (Section~\ref{sec:implementation_details_integrator}). Note that while this integrator is a coarse approximation of the underlying physical model described in Section~\ref{sec:method}, it is less important in this application, since the parameters for the kernel primitives are optimized to match the input data, thus compensating for the approximation. 
This simplification improves performance by dropping support of segment by segment integration. Since we reconstruct scenes with surface-like features, we use both Gaussian and Epanechnikov kernels in this application.
Similar to previous works~\cite{kerbl20233Dgaussians} we use a spherical harmonics expansion to model the anisotropic emission per primitive $\sRedRadE_i(\wo)$. For a detailed view on our optimization pipeline, including initialization, pruning and cloning strategies, and hyperparameters used, we refer to the supplementary material. As opposed to 3DGS, we use the same parameters in all our experiments, for both synthetic and real datasets. Fine-tuning per dataset could substantially improve the presented results.

\begin{figure*}
    \centering    
    \includegraphics[width=\textwidth]{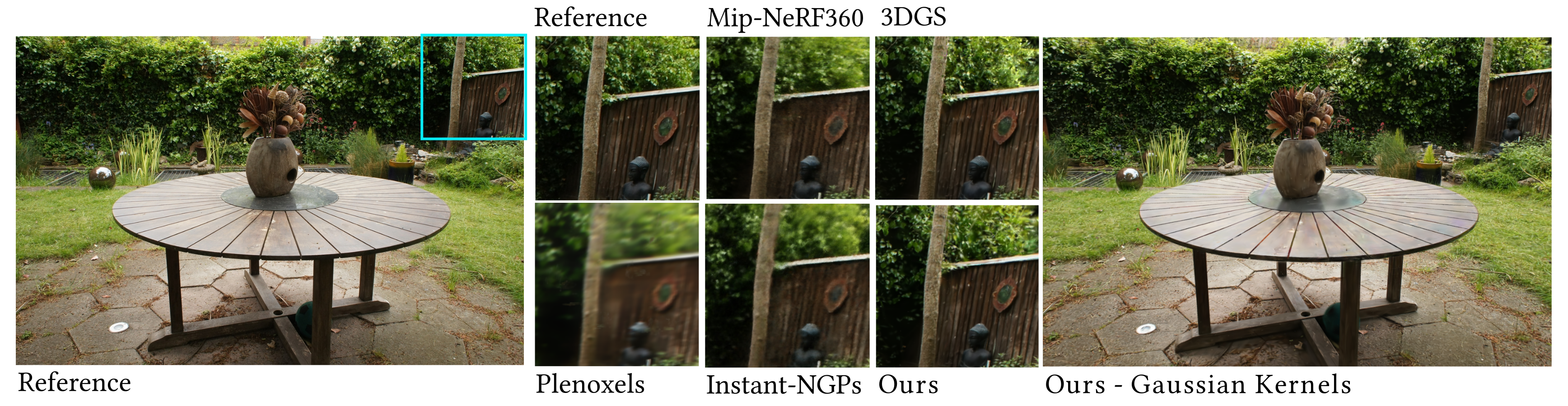}
    \includegraphics[width=\textwidth]{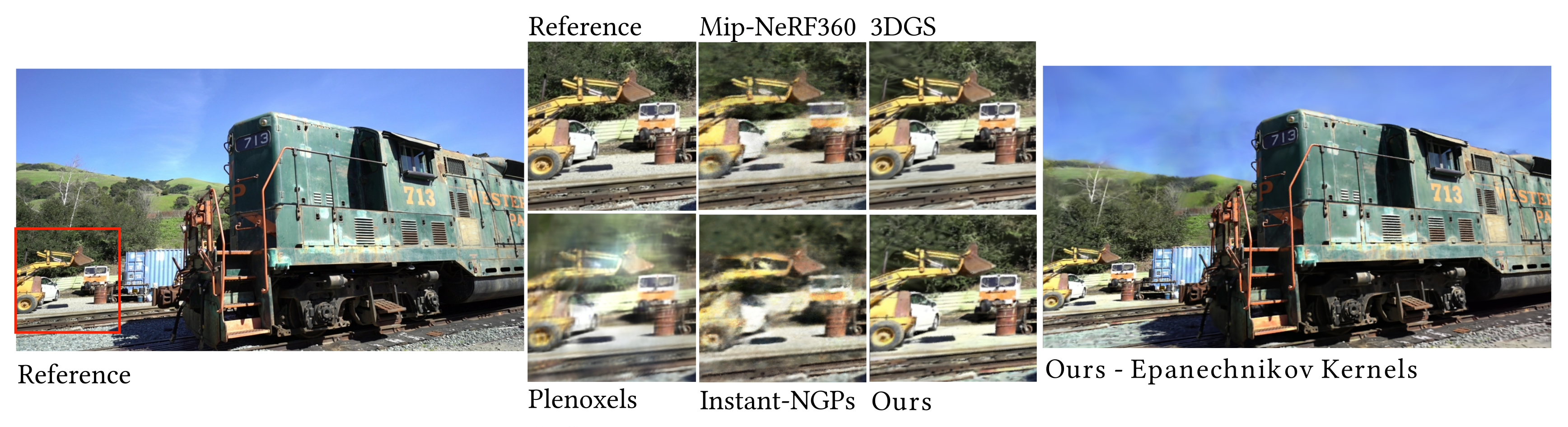}
    \includegraphics[width=\textwidth]{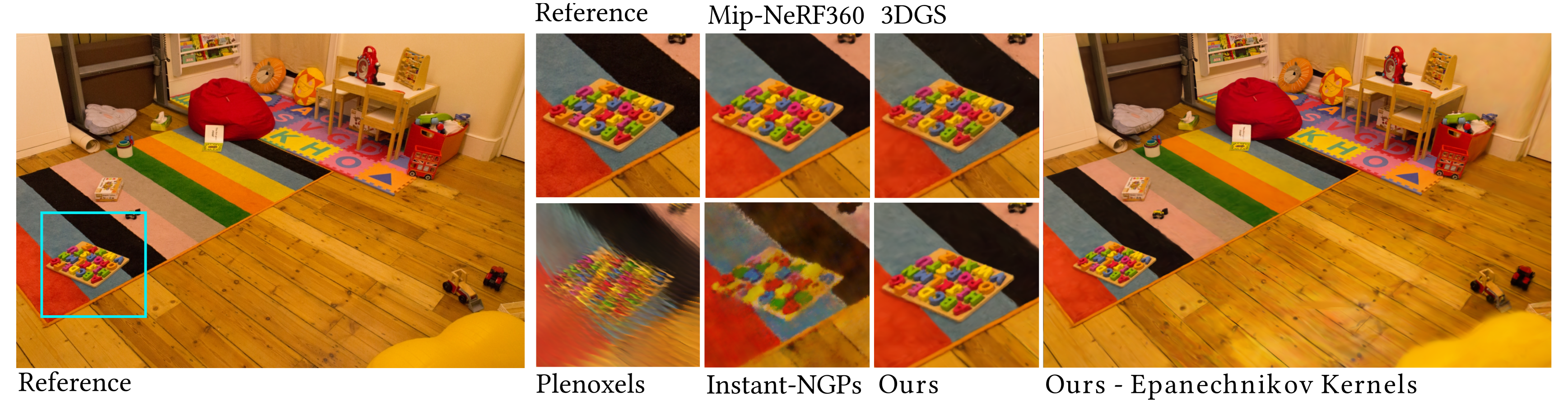}
    \includegraphics[width=\textwidth]{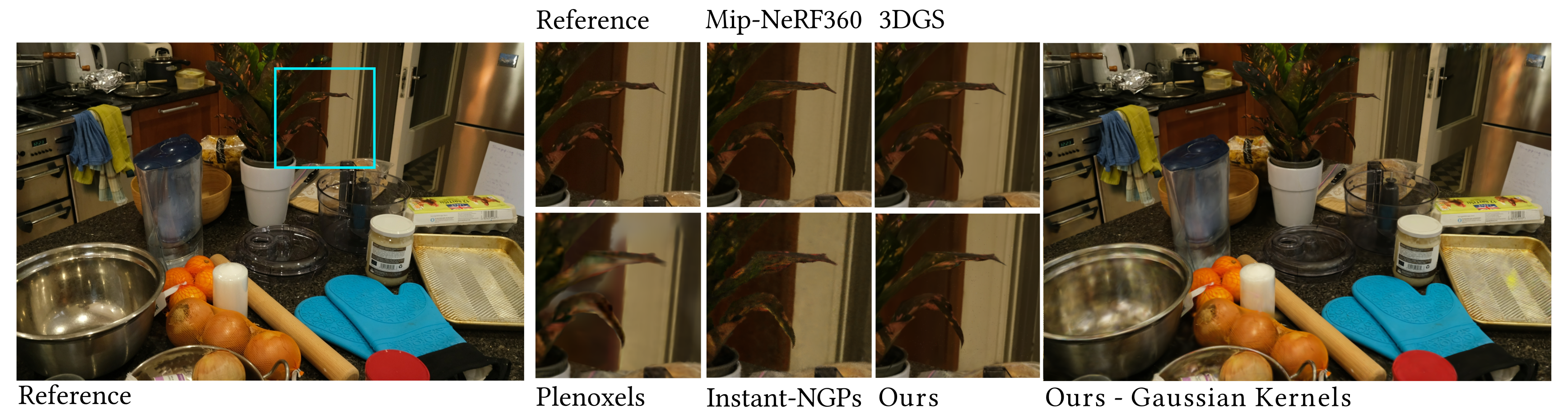}
    \caption{Qualitative comparison against previous radiance field rendering methods, including Plenoxels~\cite{yu21plenoxels}, Instant NGPs~\cite{mueller2022instant} and Mip-NeRF360~\cite{barron2022mipnerf360}, and the rasterized 3DGS~\cite{kerbl20233Dgaussians}, in the real datasets \emph{garden},  \emph{train}, \emph{playroom} and \emph{counter} (Tanks and Temples~\cite{Knapitsch2017} and MipNeRF360 datasets~\cite{barron2022mipnerf360}). Our method offers competitive quality with respect to the state of the art, and it stands as the fastest ray-traced solution (Table~\ref{tab:REVISION_rf_overview}). Both our tested kernels perform similarly in terms of quality in real scenes, while Epanechnikov kernels are substantially faster. Complete quantitative evaluations can be found in Appendix, as well as expanded images of the insets presented in this figure}.
    \label{fig:REVISION_results}
    \Description[<short description>]{<long description>}
\end{figure*}

\subsubsection{Results}
\begin{table}
\footnotesize
\caption{Quantitative quality comparison against competing methods on the real MipNeRF-360~\cite{barron2022mipnerf360} and Tanks~\cite{Knapitsch2017} datasets, and on the synthetic Blender dataset~\cite{mildenhall2020}. Average values for each metric and dataset are reported. Individual results and average execution times can be found in Supplemental.}
\label{tb:comparison}

\begin{tabular}{lcccccc}
                                     & \multicolumn{2}{c}{\textbf{MipNeRF-360}} & \multicolumn{2}{c}{\textbf{Tanks}} & \multicolumn{2}{c}{\textbf{NeRF-Blender}} \\
\textbf{Method}                      & \textbf{PSNR}$\uparrow$           & \textbf{SSIM}$\uparrow$           & \textbf{PSNR}$\uparrow$                 & \textbf{SSIM}$\uparrow$                 & \textbf{PSNR}$\uparrow$                 & \textbf{SSIM}$\uparrow$                 \\ \hline
NeRF                                 & 21.76                   & 0.455                  & 21.76                        & 0.455                        & 32.54                        & 0.961                        \\
Plenoxels                            & 21.91                   & 0.496                  & 23.22                        & 0.774                        & 34.10                        & 0.975                        \\
iNGP-Base                            & 22.19                   & 0.491                  & 23.26                        & 0.779                        & 35.64                        & 0.981                        \\
Mip-NeRF 360                  & 24.31                   & 0.685                  & 24.91                        & 0.857                        & 36.10                        & 0.980                        \\
3D Gaussian Splatting         & 25.25                   & 0.771                  & 25.19                        & 0.879                        & 36.073                       & 0.982                        \\
\textbf{Ours - Gaussian}     & 22.07                   & 0.536                  & 23.93                        & 0.850                        & 32.11                        & 0.957                        \\
\textbf{Ours - Epanechnikov} & 21.56                   & 0.515                  & 23.98                        & 0.845                        & 33.60                        & 0.972                       
\end{tabular}
\end{table}
\begin{table*}
  \caption{\textbf{Qualitative comparison of our method with other popular radiance-field (RF) approaches}. Reported performance metrics are averages over the tested real datasets, using quality (lower bound) and performance (upper bound) presets in our case. The complete set of results can be found in Supplemental. All timings are reported on similar setups using an RTX A6000, with timings of previous works taken from~\cite{kerbl20233Dgaussians}.}
  \label{tab:REVISION_rf_overview}

  \centering
  \setlength\tabcolsep{3pt}
  \resizebox{\textwidth}{!}{
  \begin{tabular}{l c c c c c c c}
    \toprule \multirow{2}{*}{Method} & NeRF & Plenoxels  & iNGP & Mip-NeRF 360 & 3DGS  & \textbf{Ours - Gaussian} & \textbf{Ours - Epanechnikov} \\ 
    & ~\cite{mildenhall2020} & ~\cite{yu21plenoxels} & ~\cite{mueller2022instant} & ~\cite{barron2022mipnerf360} & ~\cite{kerbl20233Dgaussians} \\
    \midrule
     Optimization time (RF)         & $\sim$12-24h & $\sim$0.1-0.5h & $\sim$2-8m & $\sim$48h & $\sim$0.3-0.5h & $\sim$2h-16h & 1h-10h \\
     Rendering speed (RF, real datasets)      & $\sim$0.05-0.1 FPS & $\sim$6-13 FPS & $\sim$3-17 FPS & $\sim$0.06-0.14 FPS & $\sim$134-197 FPS & $\sim$28-39 FPS& $\sim$42-53 FPS \\
     Memory footprint (RF)         & $\sim$5-10MB & $\sim$1-2.7GB & $\sim$13-48MB & $\sim$5-10MB & $\sim$100-730MB & $\sim$100-900MB & $\sim$100-900MB \\
     Rendering framework        & Ray-tracing & Ray-tracing & Ray-tracing & Ray-tracing & Rasterization & Ray-tracing & Ray-tracing \\
     Complex camera models      & Yes & Yes & Yes & Yes & No & Yes & Yes \\
    \bottomrule
  \end{tabular}}
  \vspace{2mm}
\end{table*}
Figure~\ref{fig:REVISION_results} shows qualitative comparisons against previous works~\cite{mildenhall2020,yu21plenoxels,mueller2022instant,kerbl20233Dgaussians, barron2022mipnerf360}. Quantitative average quality metrics can be found on Table~\ref{tb:comparison}, while Table~\ref{tab:REVISION_rf_overview} gives an overview of where our method stands with respect to previous work in terms of compression, rendering times, optimization times and other metrics. A complete quantitative evaluation can be found in Supplemental. 

Our results suggest comparable performance in terms of image quality to previous ray-traced radiance field solutions (e.g., instant NGP~\cite{mueller2022instant}), while falling behind the state-of-the-art, rasterized 3DGS~\cite{kerbl20233Dgaussians}. However, our method is faster than previous radiance field ray-traced works, achieving framerates of $\sim$40-50 FPS on complex scenes on average using quality presets. In many cases, the Epanechnikov kernel not only outperforms Gaussian kernels in rendering speed, but also in visual quality.

\begin{figure}
    \centering
    \includegraphics[width=\columnwidth]{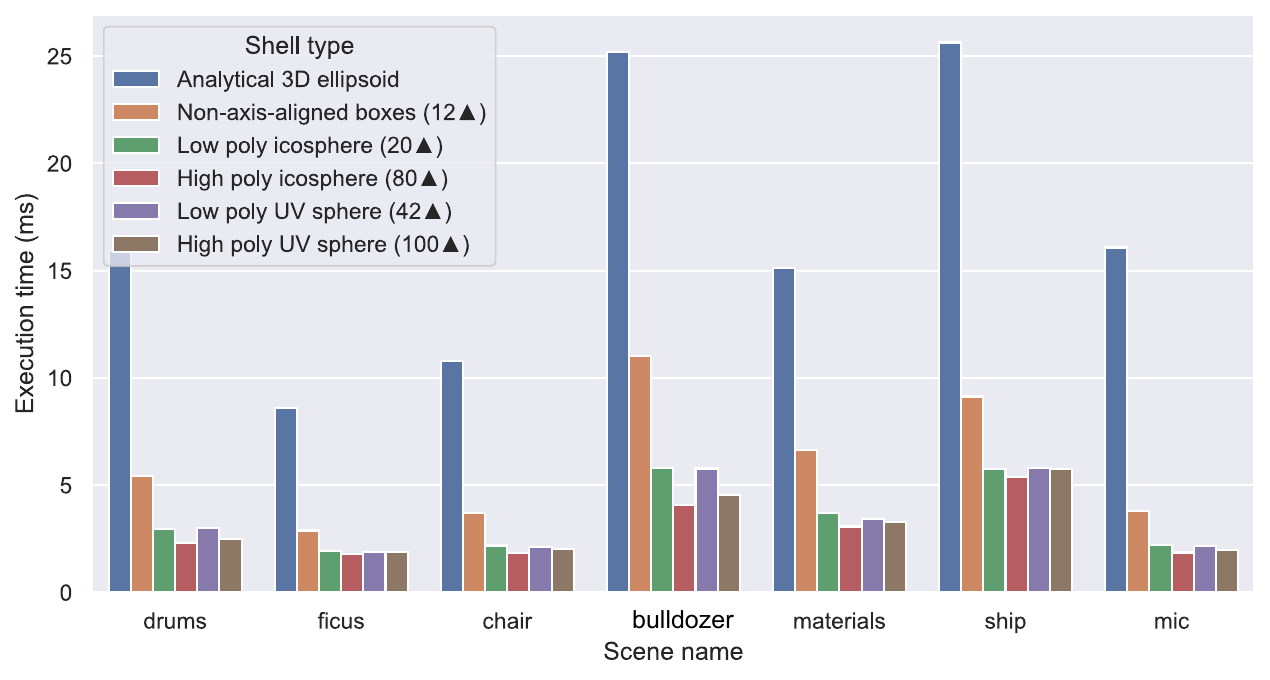}
    \caption{Rendering times using different shells across synthetic dataset scenes optimized with our method. Triangle-mesh-based shells excel at performance due to hardware ray-triangle intersection support and a more efficient traversal, at the cost of some extra memory per primitive. On average, we got a speedup of X4.96 by using triangle mesh icospheres vs analytical 3D ellipsoids. }
    \label{fig:shell_ablation}
    \Description[<short description>]{<long description>}
\end{figure}

\paragraph{Ablation on the choice of shell}
We use analytic ellipsoids as shells during the optimization due to their memory compactness. However, as discussed in Section~\ref{sec:method}, we propose several alternatives when rendering (forward or inverse) with bigger memory budgets. In Figure~\ref{fig:shell_ablation} we include an ablation over the synthetic datasets we optimized, where we show up to 4.96 times faster rendering speed when using alternative triangle-mesh based shells.

\begin{figure}
    \centering
    \includegraphics[width=\columnwidth]{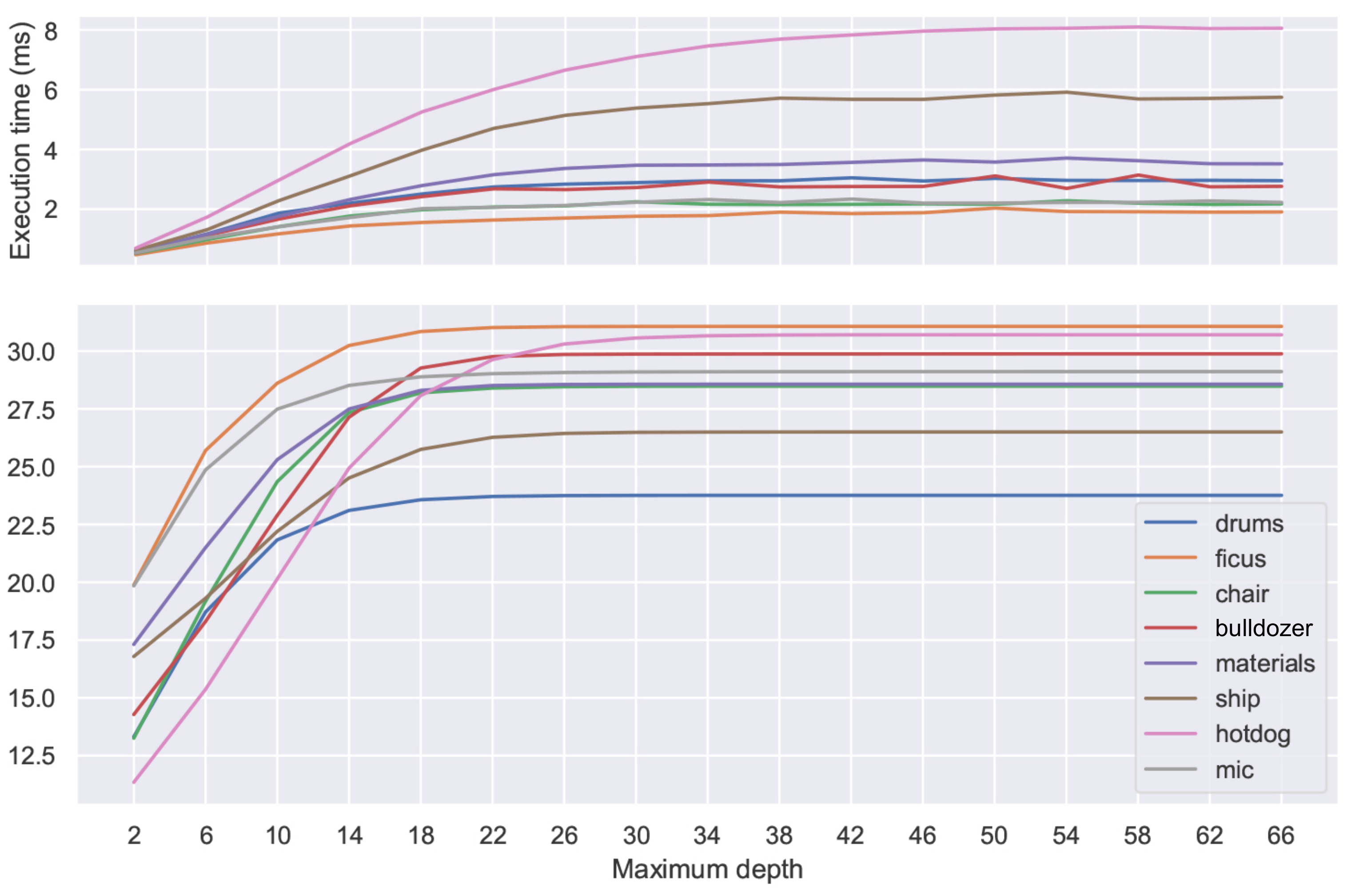}
    \caption{Effect of maximum depth (number of Gaussians integrated in the ray), as a function of PSNR quality and rendering time, on synthetic datasets. At a maximum depth of 14 Gaussians/ray, we averaged 2.36ms per frame (434.78 FPS) across all datasets, while achieving a good balance between quality and performance. With a more aggressive setting, at 10 Gaussians/ray, we measured an average execution time of 1.82ms (549.45 FPS).}
    \label{fig:rendering_times_performance_runs}
    \Description[<short description>]{<long description>}
\end{figure}

\paragraph{Ablation on Maximum Primitive Depth}
We can trade off quality for performance by limiting maximum primitive depth. Since 3DGS sorts the primitives before splatting, the potential performance gain by early integration termination is limited. In contrast, using ray-tracing can provide significant speedups in rendering times by terminating recursion early, as shown in Figure~\ref{fig:rendering_times_performance_runs}.

\subsubsection{Comparison with 3DGS}
\begin{figure}
    \centering
    \includegraphics[width=0.8\columnwidth]{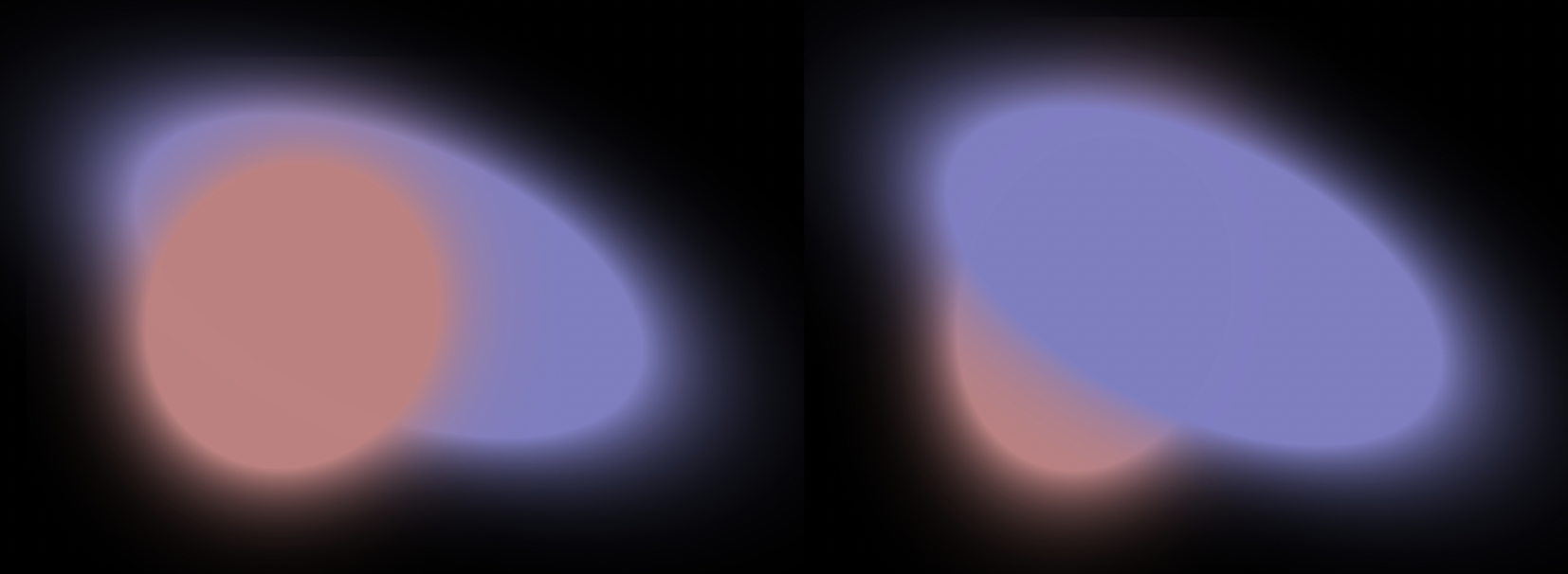}
    \caption{Splatting a mean-sorted stack of Gaussians creates noticeable artifacts. Rotating 3DGS Gaussians can "jump" order sharply depending on viewpoint, which is normally compensated through the optimization process by adding more Gaussians.}
    \label{fig:sorting_artifacts}
    \Description[<short description>]{<long description>}
\end{figure}

While slower than the rasterization-based baseline 3DGS, our ray-traced volumetric model, derived from the RTE introduces several advantages, including:

\paragraph{Accurate Ordering} Due to the inherent order obtained through recursive ray-tracing, we do not need to sort our primitives, and our order is exact, taking the anisotropic shape of our primitives into account. This avoids potential ordering artifacts observed in 3DGS, where primitives are sorted solely on their means and transformed to billboards when splatted, which may cause popping artifacts (Figure~\ref{fig:sorting_artifacts}) and inconsistent ordering among different views.

\begin{figure}
    \centering
    \includegraphics[width=\columnwidth]{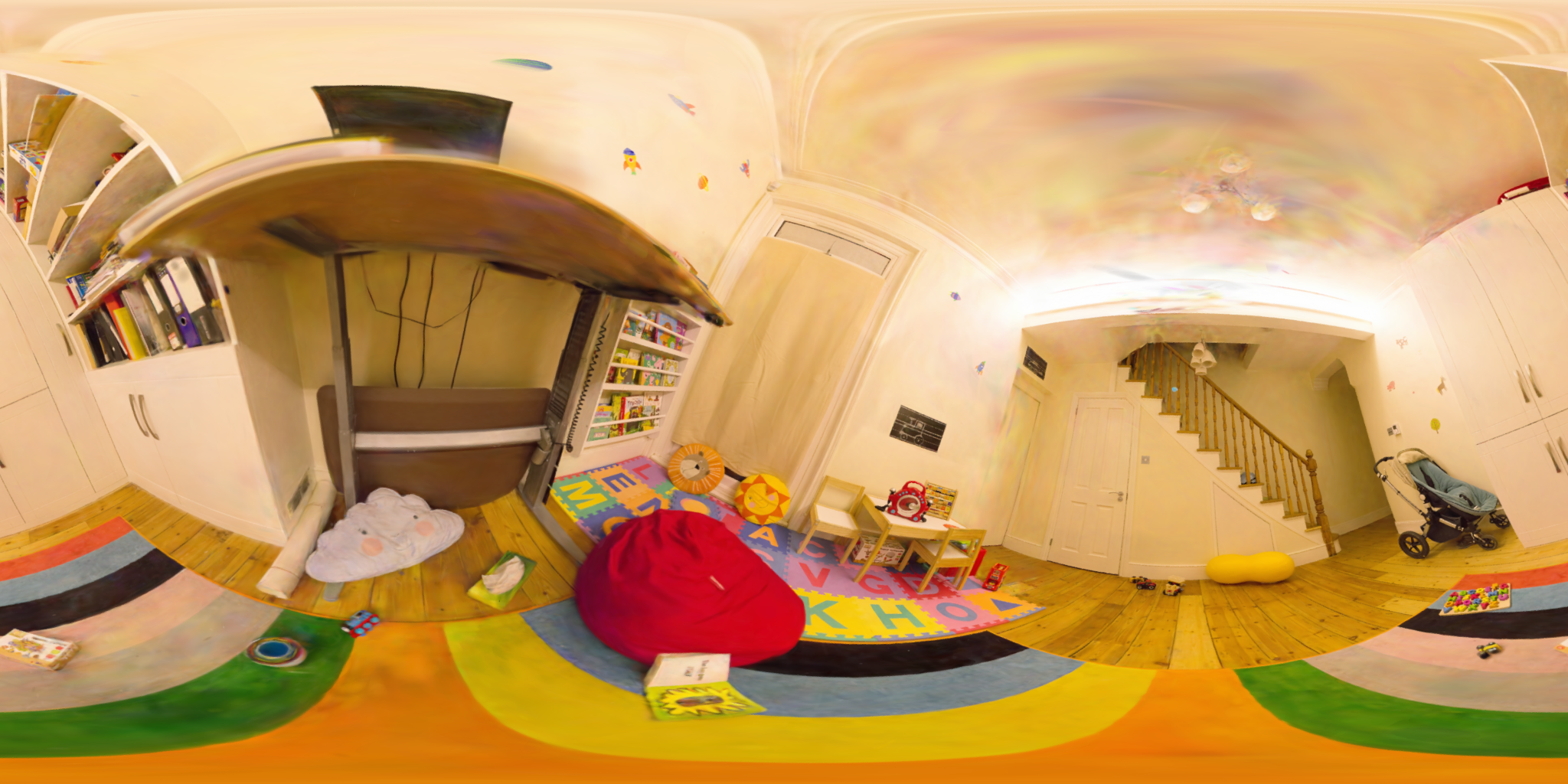}
    \caption{Complex camera models in radiance fields. We render the \emph{playroom} dataset (MipNeRF360 datasets~\cite{barron2022mipnerf360}) using a 360-degree camera. Our ray-tracing-based framework allows us to use arbitrarily complex camera models, as opposed to 3DGS. Artifacts on the ceiling and under the table are due to undersampling in the training data.}
    \label{fig:360degRF}
    \Description[<short description>]{<long description>}
\end{figure}

\paragraph{Ray tracing}
Among the many advantages of working in a ray-tracing framework compared to rasterization are the possibility of employing complex camera models without any perspective corrections for both reconstruction and rendering. We showcase the latter by rendering one of our RF optimized scenes with a 360-degree camera in Figure~\ref{fig:360degRF}. It is also straightforward to support better foveated and wide-angle rendering in VR applications and further accelerate path-tracing pipelines that rely on precomputed radiance emission to model complex light sources~\cite{zhu2021neural, condor22}.
By leveraging prior work on early ray-termination, we also obtain a simple way of controlling rendering speed at the cost of some quality. A physically-based ray-traced formulation further enables future extensions towards relighting, global illumination, soft shadows, etc.

\section{Comparison between kernels}
\label{sec:kernelcomp}
Finally, we assess the differences between the two proposed kernels. Gaussian kernels have properties that could be desirable in some situations (e.g. the convolution of Gaussians is Gaussian; convolving Gaussian kernels over the GMMs could be used to simulate animation, LoD, and motion blur~\cite{Leimkuhler2018} efficiently). In contrast, Epanechnikov kernels have limited support, which creates more compact primitives, resulting in more efficient acceleration structures and a sparser representation. 
We showcase these differences on a couple of experiments.

\begin{figure}
    \centering
    \includegraphics[width=\columnwidth]{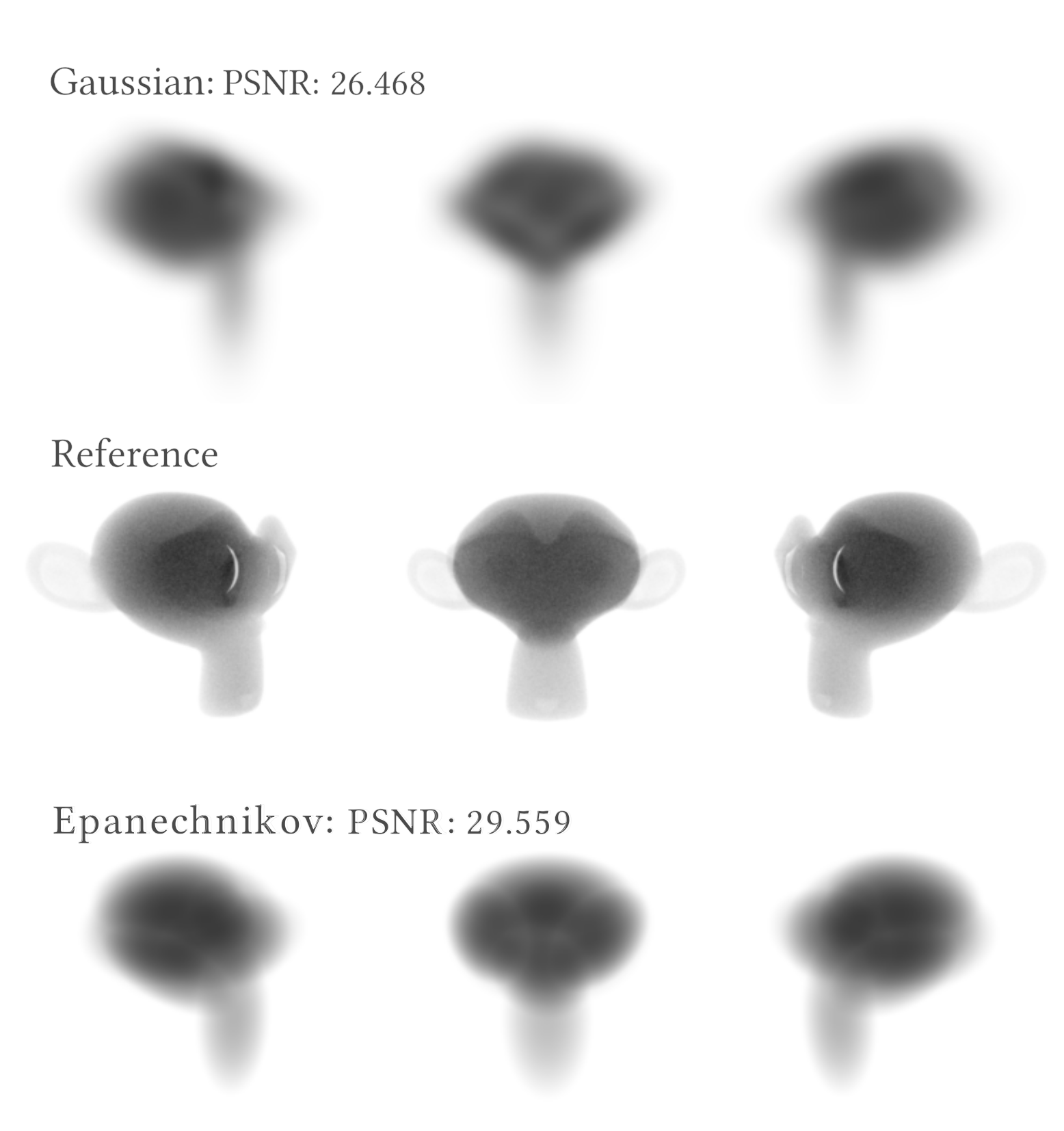}
    \caption{Example toy tomography of a mesh-bounded absorbing medium. We render 8 different views around the volume. For both kernels, we initialize on the same 8 randomly sampled primitives and optimize this fixed set until convergence (1-2 minutes). The Epanechnikov kernel performs substantially better in this controlled example; its faster decay seems to better model sharp edges. We observed a similar behaviour in our (uncontrolled) experiments with radiance field optimizations in Section~\ref{sec:radiancefields}.}
    \label{fig:kernel_comparisons}
    \Description[<short description>]{<long description>}
\end{figure}

\paragraph{Regression quality comparison}
We set up a simple inverse tomography experiment, where we optimize an heterogeneous absorbing medium and with an 8-primitives mixture model. The absorbing media is surrounded by a constant environmental emitter. The results of this experiment can be seen in Figure~\ref{fig:kernel_comparisons}. The sharp decay and smaller footprint of Epanechnikov kernel results in much sharper fits. This behavior makes them a potentially better candidate for hard surfaces and high-frequency detail.

\begin{figure}
    \centering
    \includegraphics[width=\columnwidth]{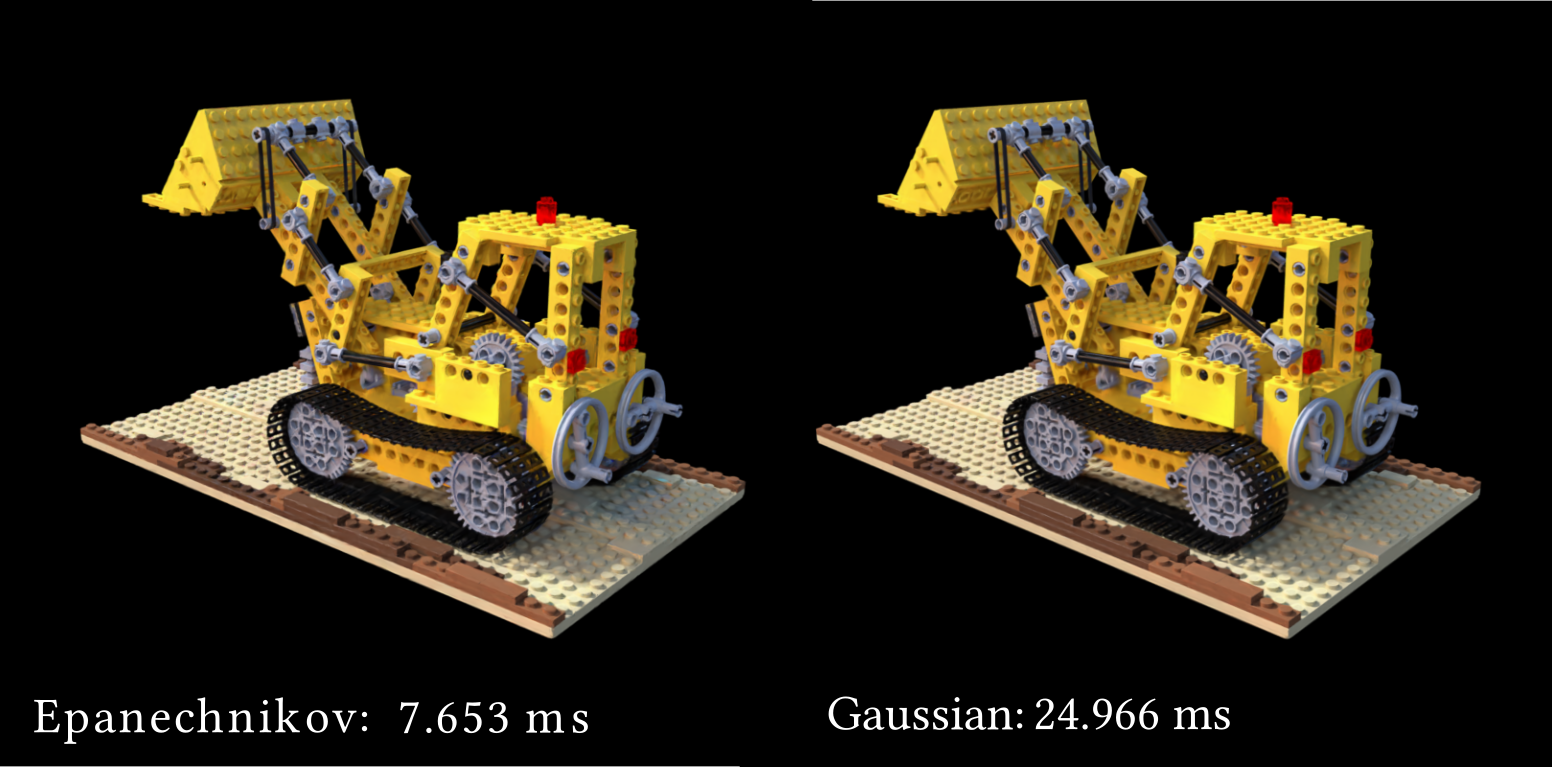}
    \caption{Speed comparison between Gaussian and Epanechnikov kernels. For virtually the same radiance field (PSNR between the different mixture renders of 38.05 dB), the Epanechnikov kernel is 3.26X faster than the Gaussian one (1spp, average over 6 views, analytic shells).}
    \label{fig:radiance_field_comp_epanechnikov}
    \Description[<short description>]{<long description>}
\end{figure}

\paragraph{Rendering speed comparison}
In Figure~\ref{fig:radiance_field_comp_epanechnikov} we compare the rendering performance from two mixtures (Gaussian vs Epanechnikov) on a more complex radiance field example. We use our radiance field pipeline (Section~\ref{sec:radiancefields}), and train a Gaussian mixture model of the \emph{Bulldozer} asset. Then, we swap the Gaussian kernel with the Epanechnikov, and optimize solely the scale for a few more iterations (using the Gaussian renders as a reference, not the original images). Producing virtually the same images, the Epanechnikov-based mixtures are much faster mainly due to the compactness of the primitive shells. The main performance bottleneck of our approach is ray traversal in a potentially degenerate BVH due to large overlapping kernels, which substantially improves with compaction. The difference can be even greater on large-scale scenes, where a single large primitive (e.g., in the sky) can substantially decrease overall performance.

\section{Discussion}
\label{sec:futurework}
We have presented a general framework for model scattering and emissive media using volumetric primitives based on 3D kernels. Together with an efficient implementation in different integrators (and their adjoints), we have shown a variety of applications including physically-based forward and inverse rendering of media, as well as radiance field rendering. 
Our volumetric primitives are flexible, compact and very efficient to render. We have shown the versatility of our approach in using different kernels, and the potential benefits of extending beyond the classic Gaussian kernel in rendering tasks. We have cleared the way for the introduction of other kinds of kernels in the future, and expect the method to be of great interest in a media production environment progressively more interested in real-time solutions for path tracing and physically-based rendering.

\paragraph{Limitations \& Future Work}
Our work presents several limitations and opportunities for future work. 
First of all, exploring how to generalize our radiative framework to anisotropic or correlated media will increase the range of scenes our model can support, including volume-based representations of scenes with solid surfaces. Exploring how a primitive-based general media can be used for level-of-detail of complex appearance is also a promising avenue of future work, compared to voxel grid-based approaches~\cite{loubet2017,vicini2021non}.

Our inverse optimization pipeline demonstrates the potential of our method for inverse rendering tasks, but we acknowledge that more work is needed to optimize for more complex volumetric media. This would require a more thorough analysis on the initialization, losses and optimization strategies, specially as we increase the complexity of the underlying light transport theory. This would significantly improve how our assets are modeled, which for now are limited to density-only heterogeneities.

As for the kernels proposed, there is a myriad of potentially good candidates, and different applications might benefit from different kernels. One particularly interesting avenue of future work would be to use parametric kernels, which could potentially bring higher flexibility and expressive power per primitive, reducing the number of primitives required for modelling an asset.

Another limitation of our inverse pipelines is the comparatively longer training times when compared to other alternatives. A large contributor to this is the JIT-compilation of the scene, which is re-run for every new iteration. This can be substantially improved by re-using compiled code across iterations, which is merely a technical limitation of DrJIT and could be lifted in the near future.

In terms of absolute rendering quality, our radiance fields are slightly lacking when compared with current state of the art~\cite{kerbl20233Dgaussians}. Still, we perform similarly to previous ray-tracing based solutions while offering much faster rendering and gracefully-degrading quality-performance tools. Furthermore, a physically-based formulation for modelling and rendering can bring many advantages over a rasterized framework (i.e. shadows, scattering, complex camera models, integration into physically-based rendereres, etc). We believe more development on the optimization pipeline can easily bring it up to par with 3DGS in quality as well.

\section*{Acknowledgements}
\label{sec:acknowledgements}
We would like to thank Christophe Hery and Olivier Maury for their continued support throughout this project and proof-reading. Jorge Condor and Piotr Didyk have been supported by the Swiss National Science Foundation (SNSF, Grant 200502) and an academic gift from Meta. 

\bibliographystyle{ACM-Reference-Format}
\bibliography{src_bib}


\begin{thebibliography}{94}


\ifx \showCODEN    \undefined \def \showCODEN     #1{\unskip}     \fi
\ifx \showDOI      \undefined \def \showDOI       #1{#1}\fi
\ifx \showISBNx    \undefined \def \showISBNx     #1{\unskip}     \fi
\ifx \showISBNxiii \undefined \def \showISBNxiii  #1{\unskip}     \fi
\ifx \showISSN     \undefined \def \showISSN      #1{\unskip}     \fi
\ifx \showLCCN     \undefined \def \showLCCN      #1{\unskip}     \fi
\ifx \shownote     \undefined \def \shownote      #1{#1}          \fi
\ifx \showarticletitle \undefined \def \showarticletitle #1{#1}   \fi
\ifx \showURL      \undefined \def \showURL       {\relax}        \fi
\providecommand\bibfield[2]{#2}
\providecommand\bibinfo[2]{#2}
\providecommand\natexlab[1]{#1}
\providecommand\showeprint[2][]{arXiv:#2}

\bibitem[Aliaga et~al\mbox{.}(2017)]%
        {aliaga17}
\bibfield{author}{\bibinfo{person}{Carlos Aliaga}, \bibinfo{person}{Carlos Castillo}, \bibinfo{person}{Diego Gutiérrez}, \bibinfo{person}{Miguel~A. Otaduy}, \bibinfo{person}{Jorge Lopez-Moreno}, {and} \bibinfo{person}{Adrián Jarabo}.} \bibinfo{year}{2017}\natexlab{}.
\newblock \showarticletitle{An Appearance Model for Textile Fibers}.
\newblock \bibinfo{journal}{\emph{Computer Graphics Forum}} \bibinfo{volume}{36}, \bibinfo{number}{4} (\bibinfo{year}{2017}).
\newblock


\bibitem[Ament et~al\mbox{.}(2014)]%
        {ament14refractive}
\bibfield{author}{\bibinfo{person}{Marco Ament}, \bibinfo{person}{Christoph Bergmann}, {and} \bibinfo{person}{Daniel Weiskopf}.} \bibinfo{year}{2014}\natexlab{}.
\newblock \showarticletitle{Refractive radiative transfer equation}.
\newblock \bibinfo{journal}{\emph{ACM Trans. Graph.}} \bibinfo{volume}{33}, \bibinfo{number}{2} (\bibinfo{year}{2014}).
\newblock


\bibitem[Barron et~al\mbox{.}(2021)]%
        {mipnerf}
\bibfield{author}{\bibinfo{person}{Jonathan~T. Barron}, \bibinfo{person}{Ben Mildenhall}, \bibinfo{person}{Matthew Tancik}, \bibinfo{person}{Peter Hedman}, \bibinfo{person}{Ricardo Martin-Brualla}, {and} \bibinfo{person}{Pratul~P. Srinivasan}.} \bibinfo{year}{2021}\natexlab{}.
\newblock \showarticletitle{Mip-NeRF: A Multiscale Representation for Anti-Aliasing Neural Radiance Fields}. In \bibinfo{booktitle}{\emph{Proceedings of ICCV}}.
\newblock


\bibitem[Barron et~al\mbox{.}(2022)]%
        {barron2022mipnerf360}
\bibfield{author}{\bibinfo{person}{Jonathan~T. Barron}, \bibinfo{person}{Ben Mildenhall}, \bibinfo{person}{Dor Verbin}, \bibinfo{person}{Pratul~P. Srinivasan}, {and} \bibinfo{person}{Peter Hedman}.} \bibinfo{year}{2022}\natexlab{}.
\newblock \showarticletitle{Mip-NeRF 360: Unbounded Anti-Aliased Neural Radiance Fields}. In \bibinfo{booktitle}{\emph{Proceedings of CVPR}}.
\newblock


\bibitem[Bitterli et~al\mbox{.}(2018)]%
        {bitterli2018radiative}
\bibfield{author}{\bibinfo{person}{Benedikt Bitterli}, \bibinfo{person}{Srinath Ravichandran}, \bibinfo{person}{Thomas M{\"u}ller}, \bibinfo{person}{Magnus Wrenninge}, \bibinfo{person}{Jan Nov{\'a}k}, \bibinfo{person}{Steve Marschner}, {and} \bibinfo{person}{Wojciech Jarosz}.} \bibinfo{year}{2018}\natexlab{}.
\newblock \showarticletitle{A radiative transfer framework for non-exponential media}.
\newblock \bibinfo{journal}{\emph{ACM Trans. Graph.}} \bibinfo{volume}{37}, \bibinfo{number}{6} (\bibinfo{year}{2018}).
\newblock


\bibitem[Bo{\v{z}}i{\v{c}} et~al\mbox{.}(2022)]%
        {bovzivc2022neu=ral}
\bibfield{author}{\bibinfo{person}{Alja{\v{z}} Bo{\v{z}}i{\v{c}}}, \bibinfo{person}{Denis Gladkov}, \bibinfo{person}{Luke Doukakis}, {and} \bibinfo{person}{Christoph Lassner}.} \bibinfo{year}{2022}\natexlab{}.
\newblock \showarticletitle{Neural Assets: Volumetric Object Capture and Rendering for Interactive Environments}.
\newblock \bibinfo{journal}{\emph{arXiv preprint arXiv:2212.06125}} (\bibinfo{year}{2022}).
\newblock


\bibitem[Brown and Martin(2003)]%
        {Brown03}
\bibfield{author}{\bibinfo{person}{Forrest~B Brown} {and} \bibinfo{person}{William~R Martin}.} \bibinfo{year}{2003}\natexlab{}.
\newblock \showarticletitle{Direct sampling of Monte Carlo flight paths in media with continuously varying cross-sections}. In \bibinfo{booktitle}{\emph{Proc. ANS Mathematics \& Computation Topical Meeting}}, Vol.~\bibinfo{volume}{2}.
\newblock


\bibitem[Burley et~al\mbox{.}(2018)]%
        {hyperion}
\bibfield{author}{\bibinfo{person}{Brent Burley}, \bibinfo{person}{David Adler}, \bibinfo{person}{Matt Jen-Yuan Chiang}, \bibinfo{person}{Hank Driskill}, \bibinfo{person}{Ralf Habel}, \bibinfo{person}{Patrick Kelly}, \bibinfo{person}{Peter Kutz}, \bibinfo{person}{Yining~Karl Li}, {and} \bibinfo{person}{Daniel Teece}.} \bibinfo{year}{2018}\natexlab{}.
\newblock \showarticletitle{The Design and Evolution of Disney’s Hyperion Renderer}.
\newblock \bibinfo{journal}{\emph{ACM Trans. Graph.}} \bibinfo{volume}{37}, \bibinfo{number}{3} (\bibinfo{year}{2018}).
\newblock


\bibitem[Carter et~al\mbox{.}(1972)]%
        {carter1972monte}
\bibfield{author}{\bibinfo{person}{LL Carter}, \bibinfo{person}{ED Cashwell}, {and} \bibinfo{person}{WM Taylor}.} \bibinfo{year}{1972}\natexlab{}.
\newblock \showarticletitle{Monte Carlo sampling with continuously varying cross sections along flight paths}.
\newblock \bibinfo{journal}{\emph{Nuclear science and engineering}} \bibinfo{volume}{48}, \bibinfo{number}{4} (\bibinfo{year}{1972}).
\newblock


\bibitem[Chandrasekhar(1960)]%
        {chandrasekhar1960radiative}
\bibfield{author}{\bibinfo{person}{Subrahmanyan Chandrasekhar}.} \bibinfo{year}{1960}\natexlab{}.
\newblock \bibinfo{booktitle}{\emph{Radiative Transfer}}.
\newblock \bibinfo{publisher}{Courier Corporation}.
\newblock


\bibitem[Chen et~al\mbox{.}(2023)]%
        {chen2022mobilenerf}
\bibfield{author}{\bibinfo{person}{Zhiqin Chen}, \bibinfo{person}{Thomas Funkhouser}, \bibinfo{person}{Peter Hedman}, {and} \bibinfo{person}{Andrea Tagliasacchi}.} \bibinfo{year}{2023}\natexlab{}.
\newblock \showarticletitle{Mobilenerf: Exploiting the polygon rasterization pipeline for efficient neural field rendering on mobile architectures}. In \bibinfo{booktitle}{\emph{Proceedings of CVPR}}.
\newblock


\bibitem[Christensen et~al\mbox{.}(2018)]%
        {renderman}
\bibfield{author}{\bibinfo{person}{Per Christensen}, \bibinfo{person}{Julian Fong}, \bibinfo{person}{Jonathan Shade}, \bibinfo{person}{Wayne Wooten}, \bibinfo{person}{Brenden Schubert}, \bibinfo{person}{Andrew Kensler}, \bibinfo{person}{Stephen Friedman}, \bibinfo{person}{Charlie Kilpatrick}, \bibinfo{person}{Cliff Ramshaw}, \bibinfo{person}{Marc Bannister}, \bibinfo{person}{Brenton Rayner}, \bibinfo{person}{Jonathan Brouillat}, {and} \bibinfo{person}{Max Liani}.} \bibinfo{year}{2018}\natexlab{}.
\newblock \showarticletitle{RenderMan: An Advanced Path-Tracing Architecture for Movie Rendering}.
\newblock \bibinfo{journal}{\emph{ACM Trans. Graph.}} \bibinfo{volume}{37}, \bibinfo{number}{3} (\bibinfo{year}{2018}).
\newblock


\bibitem[Condor and Jarabo(2022)]%
        {condor22}
\bibfield{author}{\bibinfo{person}{Jorge Condor} {and} \bibinfo{person}{Adrián Jarabo}.} \bibinfo{year}{2022}\natexlab{}.
\newblock \showarticletitle{A Learned Radiance-Field Representation for Complex Luminaires}. In \bibinfo{booktitle}{\emph{Proceedings of EGSR}}.
\newblock


\bibitem[Condor et~al\mbox{.}(2023)]%
        {Condor2023}
\bibfield{author}{\bibinfo{person}{Jorge Condor}, \bibinfo{person}{Michal Piovar\v{c}i}, \bibinfo{person}{Bernd Bickel}, {and} \bibinfo{person}{Piotr Didyk}.} \bibinfo{year}{2023}\natexlab{}.
\newblock \showarticletitle{Gloss-aware Color Correction for 3D Printing}. In \bibinfo{booktitle}{\emph{ACM SIGGRAPH Conference Papers}}.
\newblock


\bibitem[Cramer(1978)]%
        {cramer1978application}
\bibfield{author}{\bibinfo{person}{SN Cramer}.} \bibinfo{year}{1978}\natexlab{}.
\newblock \showarticletitle{Application of the fictitious scattering radiation transport model for deep-penetration Monte Carlo calculations}.
\newblock \bibinfo{journal}{\emph{Nuclear Science and Engineering}} \bibinfo{volume}{65}, \bibinfo{number}{2} (\bibinfo{year}{1978}).
\newblock


\bibitem[Crespo et~al\mbox{.}(2021)]%
        {crespo2021primary}
\bibfield{author}{\bibinfo{person}{Miguel Crespo}, \bibinfo{person}{Adrian Jarabo}, {and} \bibinfo{person}{Adolfo Mu{\~n}oz}.} \bibinfo{year}{2021}\natexlab{}.
\newblock \showarticletitle{Primary-space adaptive control variates using piecewise-polynomial approximations}.
\newblock \bibinfo{journal}{\emph{ACM Trans. Graph.}} \bibinfo{volume}{40}, \bibinfo{number}{3} (\bibinfo{year}{2021}).
\newblock


\bibitem[Elek et~al\mbox{.}(2017)]%
        {elek17}
\bibfield{author}{\bibinfo{person}{Oskar Elek}, \bibinfo{person}{Denis Sumin}, \bibinfo{person}{Ran Zhang}, \bibinfo{person}{Tim Weyrich}, \bibinfo{person}{Karol Myszkowski}, \bibinfo{person}{Bernd Bickel}, \bibinfo{person}{Alexander Wilkie}, {and} \bibinfo{person}{Jaroslav K\v{r}iv\'{a}nek}.} \bibinfo{year}{2017}\natexlab{}.
\newblock \showarticletitle{Scattering-Aware Texture Reproduction for 3D Printing}.
\newblock \bibinfo{journal}{\emph{ACM Trans. Graph.}} \bibinfo{volume}{36}, \bibinfo{number}{6} (\bibinfo{year}{2017}).
\newblock


\bibitem[Fascione et~al\mbox{.}(2018)]%
        {manuka}
\bibfield{author}{\bibinfo{person}{Luca Fascione}, \bibinfo{person}{Johannes Hanika}, \bibinfo{person}{Mark Leone}, \bibinfo{person}{Marc Droske}, \bibinfo{person}{Jorge Schwarzhaupt}, \bibinfo{person}{Tom\'{a}\v{s} Davidovi\v{c}}, \bibinfo{person}{Andrea Weidlich}, {and} \bibinfo{person}{Johannes Meng}.} \bibinfo{year}{2018}\natexlab{}.
\newblock \showarticletitle{Manuka: A Batch-Shading Architecture for Spectral Path Tracing in Movie Production}.
\newblock \bibinfo{journal}{\emph{ACM Trans. Graph.}} \bibinfo{volume}{37}, \bibinfo{number}{3} (\bibinfo{year}{2018}).
\newblock


\bibitem[Fong et~al\mbox{.}(2017)]%
        {ProductionVolumeRendering2017}
\bibfield{author}{\bibinfo{person}{Julian Fong}, \bibinfo{person}{Magnus Wrenninge}, \bibinfo{person}{Christopher Kulla}, {and} \bibinfo{person}{Ralf Habel}.} \bibinfo{year}{2017}\natexlab{}.
\newblock \showarticletitle{Production volume rendering}. In \bibinfo{booktitle}{\emph{ACM SIGGRAPH 2017 Courses}}.
\newblock


\bibitem[Fridovich-Keil et~al\mbox{.}(2022)]%
        {yu21plenoxels}
\bibfield{author}{\bibinfo{person}{Sara Fridovich-Keil}, \bibinfo{person}{Alex Yu}, \bibinfo{person}{Matthew Tancik}, \bibinfo{person}{Qinhong Chen}, \bibinfo{person}{Benjamin Recht}, {and} \bibinfo{person}{Angjoo Kanazawa}.} \bibinfo{year}{2022}\natexlab{}.
\newblock \showarticletitle{Plenoxels: Radiance fields without neural networks}. In \bibinfo{booktitle}{\emph{Proceedings of CVPR}}.
\newblock


\bibitem[Galtier et~al\mbox{.}(2013)]%
        {galtier2013integral}
\bibfield{author}{\bibinfo{person}{Mathieu Galtier}, \bibinfo{person}{Stephane Blanco}, \bibinfo{person}{Cyril Caliot}, \bibinfo{person}{Christophe Coustet}, \bibinfo{person}{J{\'e}r{\'e}mi Dauchet}, \bibinfo{person}{Mouna El~Hafi}, \bibinfo{person}{Vincent Eymet}, \bibinfo{person}{Richard Fournier}, \bibinfo{person}{Jacques Gautrais}, \bibinfo{person}{Anais Khuong}, {et~al\mbox{.}}} \bibinfo{year}{2013}\natexlab{}.
\newblock \showarticletitle{Integral formulation of null-collision Monte Carlo algorithms}.
\newblock \bibinfo{journal}{\emph{Journal of Quantitative Spectroscopy and Radiative Transfer}}  \bibinfo{volume}{125} (\bibinfo{year}{2013}).
\newblock


\bibitem[Georgiev et~al\mbox{.}(2018)]%
        {fajardo18arnold}
\bibfield{author}{\bibinfo{person}{Iliyan Georgiev}, \bibinfo{person}{Thiago Ize}, \bibinfo{person}{Mike Farnsworth}, \bibinfo{person}{Ram\'{o}n Montoya-Vozmediano}, \bibinfo{person}{Alan King}, \bibinfo{person}{Brecht~Van Lommel}, \bibinfo{person}{Angel Jimenez}, \bibinfo{person}{Oscar Anson}, \bibinfo{person}{Shinji Ogaki}, \bibinfo{person}{Eric Johnston}, \bibinfo{person}{Adrien Herubel}, \bibinfo{person}{Declan Russell}, \bibinfo{person}{Fr\'{e}d\'{e}ric Servant}, {and} \bibinfo{person}{Marcos Fajardo}.} \bibinfo{year}{2018}\natexlab{}.
\newblock \showarticletitle{Arnold: A Brute-Force Production Path Tracer}.
\newblock \bibinfo{journal}{\emph{ACM Trans. Graph.}} \bibinfo{volume}{37}, \bibinfo{number}{3} (\bibinfo{year}{2018}).
\newblock


\bibitem[Georgiev et~al\mbox{.}(2019)]%
        {georgiev19}
\bibfield{author}{\bibinfo{person}{Iliyan Georgiev}, \bibinfo{person}{Zackary Misso}, \bibinfo{person}{Toshiya Hachisuka}, \bibinfo{person}{Derek Nowrouzezahrai}, \bibinfo{person}{Jaroslav K\v{r}iv\'{a}nek}, {and} \bibinfo{person}{Wojciech Jarosz}.} \bibinfo{year}{2019}\natexlab{}.
\newblock \showarticletitle{Integral formulations of volumetric transmittance}.
\newblock \bibinfo{journal}{\emph{ACM Trans. Graph.}} \bibinfo{volume}{38}, \bibinfo{number}{6} (\bibinfo{year}{2019}).
\newblock


\bibitem[Gkioulekas et~al\mbox{.}(2016)]%
        {Gkioulekas2016AFF}
\bibfield{author}{\bibinfo{person}{Ioannis Gkioulekas}, \bibinfo{person}{Anat Levin}, {and} \bibinfo{person}{Todd Zickler}.} \bibinfo{year}{2016}\natexlab{}.
\newblock \showarticletitle{An evaluation of computational imaging techniques for heterogeneous inverse scattering}. In \bibinfo{booktitle}{\emph{Proceedings of ECCV}}.
\newblock


\bibitem[Green et~al\mbox{.}(2006)]%
        {green2006view}
\bibfield{author}{\bibinfo{person}{Paul Green}, \bibinfo{person}{Jan Kautz}, \bibinfo{person}{Wojciech Matusik}, {and} \bibinfo{person}{Fr{\'e}do Durand}.} \bibinfo{year}{2006}\natexlab{}.
\newblock \showarticletitle{View-dependent precomputed light transport using nonlinear gaussian function approximations}. In \bibinfo{booktitle}{\emph{Proceedings of I3D}}.
\newblock


\bibitem[Hedman et~al\mbox{.}(2021)]%
        {hedman2021snerg}
\bibfield{author}{\bibinfo{person}{Peter Hedman}, \bibinfo{person}{Pratul~P Srinivasan}, \bibinfo{person}{Ben Mildenhall}, \bibinfo{person}{Jonathan~T Barron}, {and} \bibinfo{person}{Paul Debevec}.} \bibinfo{year}{2021}\natexlab{}.
\newblock \showarticletitle{Baking neural radiance fields for real-time view synthesis}. In \bibinfo{booktitle}{\emph{Proceedings of ICCV}}.
\newblock


\bibitem[Heitz et~al\mbox{.}(2015)]%
        {heitz2015sggx}
\bibfield{author}{\bibinfo{person}{Eric Heitz}, \bibinfo{person}{Jonathan Dupuy}, \bibinfo{person}{Cyril Crassin}, {and} \bibinfo{person}{Carsten Dachsbacher}.} \bibinfo{year}{2015}\natexlab{}.
\newblock \showarticletitle{The SGGX microflake distribution}.
\newblock \bibinfo{journal}{\emph{ACM Trans. Graph.}} \bibinfo{volume}{34}, \bibinfo{number}{4} (\bibinfo{year}{2015}).
\newblock


\bibitem[Jakob(2010)]%
        {jakob2010mitsuba}
\bibfield{author}{\bibinfo{person}{Wenzel Jakob}.} \bibinfo{year}{2010}\natexlab{}.
\newblock \bibinfo{title}{Mitsuba renderer}.
\newblock
\newblock


\bibitem[Jakob et~al\mbox{.}(2010)]%
        {jakob2010radiative}
\bibfield{author}{\bibinfo{person}{Wenzel Jakob}, \bibinfo{person}{Adam Arbree}, \bibinfo{person}{Jonathan~T Moon}, \bibinfo{person}{Kavita Bala}, {and} \bibinfo{person}{Steve Marschner}.} \bibinfo{year}{2010}\natexlab{}.
\newblock \showarticletitle{A radiative transfer framework for rendering materials with anisotropic structure}.
\newblock \bibinfo{journal}{\emph{ACM Trans. Graph.}} \bibinfo{volume}{29}, \bibinfo{number}{4} (\bibinfo{year}{2010}).
\newblock


\bibitem[Jakob et~al\mbox{.}(2011)]%
        {jakob2011progressive}
\bibfield{author}{\bibinfo{person}{Wenzel Jakob}, \bibinfo{person}{Christian Regg}, {and} \bibinfo{person}{Wojciech Jarosz}.} \bibinfo{year}{2011}\natexlab{}.
\newblock \showarticletitle{Progressive {{Expectation}}--{{Maximization}} for Hierarchical Volumetric Photon Mapping}.
\newblock \bibinfo{journal}{\emph{Computer Graphics Forum}} \bibinfo{volume}{30}, \bibinfo{number}{4} (\bibinfo{year}{2011}).
\newblock


\bibitem[Jakob et~al\mbox{.}(2022b)]%
        {jakob2022mitsuba3}
\bibfield{author}{\bibinfo{person}{Wenzel Jakob}, \bibinfo{person}{Sébastien Speierer}, \bibinfo{person}{Nicolas Roussel}, \bibinfo{person}{Merlin Nimier-David}, \bibinfo{person}{Delio Vicini}, \bibinfo{person}{Tizian Zeltner}, \bibinfo{person}{Baptiste Nicolet}, \bibinfo{person}{Miguel Crespo}, \bibinfo{person}{Vincent Leroy}, {and} \bibinfo{person}{Ziyi Zhang}.} \bibinfo{year}{2022}\natexlab{b}.
\newblock \bibinfo{booktitle}{\emph{Mitsuba 3 renderer}}.
\newblock
\newblock
\shownote{https://mitsuba-renderer.org}.


\bibitem[Jakob et~al\mbox{.}(2022a)]%
        {jakob2020drjit}
\bibfield{author}{\bibinfo{person}{Wenzel Jakob}, \bibinfo{person}{Sébastien Speierer}, \bibinfo{person}{Nicolas Roussel}, {and} \bibinfo{person}{Delio Vicini}.} \bibinfo{year}{2022}\natexlab{a}.
\newblock \showarticletitle{Dr.Jit: A Just-In-Time Compiler for Differentiable Rendering}.
\newblock \bibinfo{journal}{\emph{ACM Trans. Graph.}} \bibinfo{volume}{41}, \bibinfo{number}{4} (\bibinfo{year}{2022}).
\newblock


\bibitem[Jarabo et~al\mbox{.}(2018)]%
        {jarabo2018radiative}
\bibfield{author}{\bibinfo{person}{Adrian Jarabo}, \bibinfo{person}{Carlos Aliaga}, {and} \bibinfo{person}{Diego Gutierrez}.} \bibinfo{year}{2018}\natexlab{}.
\newblock \showarticletitle{A radiative transfer framework for spatially-correlated materials}.
\newblock \bibinfo{journal}{\emph{ACM Trans. Graph.}} \bibinfo{volume}{37}, \bibinfo{number}{4} (\bibinfo{year}{2018}).
\newblock


\bibitem[Kajiya(1986)]%
        {kajiya86}
\bibfield{author}{\bibinfo{person}{James~T. Kajiya}.} \bibinfo{year}{1986}\natexlab{}.
\newblock \showarticletitle{The Rendering Equation}.
\newblock \bibinfo{journal}{\emph{SIGGRAPH Comput. Graph.}} \bibinfo{volume}{20}, \bibinfo{number}{4} (\bibinfo{year}{1986}).
\newblock


\bibitem[Kallweit et~al\mbox{.}(2017)]%
        {kallweit2017}
\bibfield{author}{\bibinfo{person}{Simon Kallweit}, \bibinfo{person}{Thomas M\"{u}ller}, \bibinfo{person}{Brian Mcwilliams}, \bibinfo{person}{Markus Gross}, {and} \bibinfo{person}{Jan Nov\'{a}k}.} \bibinfo{year}{2017}\natexlab{}.
\newblock \showarticletitle{Deep Scattering: Rendering Atmospheric Clouds with Radiance-Predicting Neural Networks}.
\newblock \bibinfo{journal}{\emph{ACM Trans. Graph.}} \bibinfo{volume}{36}, \bibinfo{number}{6} (\bibinfo{year}{2017}).
\newblock


\bibitem[Kerbl et~al\mbox{.}(2023)]%
        {kerbl20233Dgaussians}
\bibfield{author}{\bibinfo{person}{Bernhard Kerbl}, \bibinfo{person}{Georgios Kopanas}, \bibinfo{person}{Thomas Leimk{\"u}hler}, {and} \bibinfo{person}{George Drettakis}.} \bibinfo{year}{2023}\natexlab{}.
\newblock \showarticletitle{3D Gaussian Splatting for Real-Time Radiance Field Rendering}.
\newblock \bibinfo{journal}{\emph{ACM Trans. Graph.}} \bibinfo{volume}{42}, \bibinfo{number}{4} (\bibinfo{year}{2023}).
\newblock


\bibitem[Kettunen et~al\mbox{.}(2021)]%
        {kettunen21}
\bibfield{author}{\bibinfo{person}{Markus Kettunen}, \bibinfo{person}{Eugene D'Eon}, \bibinfo{person}{Jacopo Pantaleoni}, {and} \bibinfo{person}{Jan Nov\'{a}k}.} \bibinfo{year}{2021}\natexlab{}.
\newblock \showarticletitle{An unbiased ray-marching transmittance estimator}.
\newblock \bibinfo{journal}{\emph{ACM Trans. Graph.}} \bibinfo{volume}{40}, \bibinfo{number}{4} (\bibinfo{year}{2021}).
\newblock


\bibitem[Khungurn et~al\mbox{.}(2015)]%
        {khungurn2015}
\bibfield{author}{\bibinfo{person}{Pramook Khungurn}, \bibinfo{person}{Daniel Schroeder}, \bibinfo{person}{Shuang Zhao}, \bibinfo{person}{Kavita Bala}, {and} \bibinfo{person}{Steve Marschner}.} \bibinfo{year}{2015}\natexlab{}.
\newblock \showarticletitle{Matching Real Fabrics with Micro-Appearance Models.}
\newblock \bibinfo{journal}{\emph{ACM Trans. Graph.}} \bibinfo{volume}{35}, \bibinfo{number}{1} (\bibinfo{year}{2015}).
\newblock


\bibitem[Kim et~al\mbox{.}(2024)]%
        {neuralvdb}
\bibfield{author}{\bibinfo{person}{Doyub Kim}, \bibinfo{person}{Minjae Lee}, {and} \bibinfo{person}{Ken Museth}.} \bibinfo{year}{2024}\natexlab{}.
\newblock \showarticletitle{NeuralVDB: High-resolution Sparse Volume Representation using Hierarchical Neural Networks}.
\newblock \bibinfo{journal}{\emph{ACM Trans. Graph.}} \bibinfo{volume}{43}, \bibinfo{number}{2} (\bibinfo{year}{2024}).
\newblock


\bibitem[Kingma and Ba(2017)]%
        {kingma2017adam}
\bibfield{author}{\bibinfo{person}{Diederik~P. Kingma} {and} \bibinfo{person}{Jimmy Ba}.} \bibinfo{year}{2017}\natexlab{}.
\newblock \bibinfo{title}{Adam: A Method for Stochastic Optimization}.
\newblock
\newblock
\showeprint[arxiv]{1412.6980}


\bibitem[Kirk(2007)]%
        {cuda}
\bibfield{author}{\bibinfo{person}{David Kirk}.} \bibinfo{year}{2007}\natexlab{}.
\newblock \showarticletitle{{NVIDIA} {CUDA} software and {GPU} parallel computing architecture}. In \bibinfo{booktitle}{\emph{Proceedings of ISMM}}. \bibinfo{numpages}{2}~pages.
\newblock


\bibitem[Knapitsch et~al\mbox{.}(2017)]%
        {Knapitsch2017}
\bibfield{author}{\bibinfo{person}{Arno Knapitsch}, \bibinfo{person}{Jaesik Park}, \bibinfo{person}{Qian-Yi Zhou}, {and} \bibinfo{person}{Vladlen Koltun}.} \bibinfo{year}{2017}\natexlab{}.
\newblock \showarticletitle{Tanks and Temples: Benchmarking Large-Scale Scene Reconstruction}.
\newblock \bibinfo{journal}{\emph{ACM Trans. Graph.}} \bibinfo{volume}{36}, \bibinfo{number}{4} (\bibinfo{year}{2017}).
\newblock


\bibitem[Knoll et~al\mbox{.}(2021)]%
        {Knoll2021}
\bibfield{author}{\bibinfo{person}{Aaron Knoll}, \bibinfo{person}{Gregory~P. Johnson}, {and} \bibinfo{person}{Johannes Meng}.} \bibinfo{year}{2021}\natexlab{}.
\newblock \showarticletitle{Path Tracing RBF Particle Volumes}. In \bibinfo{booktitle}{\emph{Ray Tracing Gems II}}, \bibfield{editor}{\bibinfo{person}{Adam Marrs}, \bibinfo{person}{Peter Shirley}, {and} \bibinfo{person}{Ingo Wald}} (Eds.). \bibinfo{publisher}{Apress}.
\newblock


\bibitem[Kutz et~al\mbox{.}(2017)]%
        {kutz2017spectral}
\bibfield{author}{\bibinfo{person}{Peter Kutz}, \bibinfo{person}{Ralf Habel}, \bibinfo{person}{Yining~Karl Li}, {and} \bibinfo{person}{Jan Nov{\'a}k}.} \bibinfo{year}{2017}\natexlab{}.
\newblock \showarticletitle{Spectral and decomposition tracking for rendering heterogeneous volumes}.
\newblock \bibinfo{journal}{\emph{ACM Trans. Graph.}} \bibinfo{volume}{36}, \bibinfo{number}{4} (\bibinfo{year}{2017}), \bibinfo{pages}{1--16}.
\newblock


\bibitem[Leimk\"uhler et~al\mbox{.}(2018)]%
        {Leimkuhler2018}
\bibfield{author}{\bibinfo{person}{Thomas Leimk\"uhler}, \bibinfo{person}{Hans-Peter Seidel}, {and} \bibinfo{person}{Tobias Ritschel}.} \bibinfo{year}{2018}\natexlab{}.
\newblock \showarticletitle{Laplacian Kernel Splatting for Efficient Depth-of-field and Motion Blur Synthesis or Reconstruction}.
\newblock \bibinfo{journal}{\emph{ACM Trans. Graph.}} \bibinfo{volume}{37}, \bibinfo{number}{4} (\bibinfo{year}{2018}).
\newblock


\bibitem[Liu et~al\mbox{.}(2021)]%
        {liu2021epan}
\bibfield{author}{\bibinfo{person}{Boning Liu}, \bibinfo{person}{Yan Zhao}, \bibinfo{person}{Xiaomeng Jiang}, {and} \bibinfo{person}{Shigang Wang}.} \bibinfo{year}{2021}\natexlab{}.
\newblock \showarticletitle{Three-dimensional Epanechnikov mixture regression in image coding}.
\newblock \bibinfo{journal}{\emph{Signal Processing}}  \bibinfo{volume}{185} (\bibinfo{year}{2021}).
\newblock


\bibitem[Lombardi et~al\mbox{.}(2021)]%
        {Lombardi21}
\bibfield{author}{\bibinfo{person}{Stephen Lombardi}, \bibinfo{person}{Tomas Simon}, \bibinfo{person}{Gabriel Schwartz}, \bibinfo{person}{Michael Zollhoefer}, \bibinfo{person}{Yaser Sheikh}, {and} \bibinfo{person}{Jason Saragih}.} \bibinfo{year}{2021}\natexlab{}.
\newblock \showarticletitle{Mixture of Volumetric Primitives for Efficient Neural Rendering}.
\newblock \bibinfo{journal}{\emph{ACM Trans. Graph.}} \bibinfo{volume}{40}, \bibinfo{number}{4} (\bibinfo{year}{2021}).
\newblock


\bibitem[Loubet and Neyret(2017)]%
        {loubet2017}
\bibfield{author}{\bibinfo{person}{Guillaume Loubet} {and} \bibinfo{person}{Fabrice Neyret}.} \bibinfo{year}{2017}\natexlab{}.
\newblock \showarticletitle{Hybrid mesh-volume LoDs for all-scale pre-filtering of complex 3D assets}.
\newblock \bibinfo{journal}{\emph{Computer Graphics Forum}} \bibinfo{volume}{36}, \bibinfo{number}{2} (\bibinfo{year}{2017}).
\newblock


\bibitem[Max(1979)]%
        {max1979atomlll}
\bibfield{author}{\bibinfo{person}{Nelson~L Max}.} \bibinfo{year}{1979}\natexlab{}.
\newblock \showarticletitle{ATOMLLL: ATOMS with shading and highlights}.
\newblock \bibinfo{journal}{\emph{ACM SIGGRAPH Computer Graphics}} \bibinfo{volume}{13}, \bibinfo{number}{2} (\bibinfo{year}{1979}).
\newblock


\bibitem[Meng et~al\mbox{.}(2015)]%
        {meng15}
\bibfield{author}{\bibinfo{person}{Johannes Meng}, \bibinfo{person}{Marios Papas}, \bibinfo{person}{Ralf Habel}, \bibinfo{person}{Carsten Dachsbacher}, \bibinfo{person}{Steve Marschner}, \bibinfo{person}{Markus Gross}, {and} \bibinfo{person}{Wojciech Jarosz}.} \bibinfo{year}{2015}\natexlab{}.
\newblock \showarticletitle{Multi-Scale Modeling and Rendering of Granular Materials}.
\newblock \bibinfo{journal}{\emph{ACM Trans. Graph.}} \bibinfo{volume}{34}, \bibinfo{number}{4} (\bibinfo{year}{2015}).
\newblock


\bibitem[Mildenhall et~al\mbox{.}(2020)]%
        {mildenhall2020}
\bibfield{author}{\bibinfo{person}{Ben Mildenhall}, \bibinfo{person}{Pratul~P. Srinivasan}, \bibinfo{person}{Matthew Tancik}, \bibinfo{person}{Jonathan~T. Barron}, \bibinfo{person}{Ravi Ramamoorthi}, {and} \bibinfo{person}{Ren Ng}.} \bibinfo{year}{2020}\natexlab{}.
\newblock \showarticletitle{{NeRF}: Representing Scenes as Neural Radiance Fields for View Synthesis}. In \bibinfo{booktitle}{\emph{Proceedings of ECCV}}.
\newblock


\bibitem[Misso et~al\mbox{.}(2023)]%
        {misso2023progressive}
\bibfield{author}{\bibinfo{person}{Zackary Misso}, \bibinfo{person}{Yining~Karl Li}, \bibinfo{person}{Brent Burley}, \bibinfo{person}{Daniel Teece}, {and} \bibinfo{person}{Wojciech Jarosz}.} \bibinfo{year}{2023}\natexlab{}.
\newblock \showarticletitle{Progressive null-tracking for volumetric rendering}. In \bibinfo{booktitle}{\emph{ACM SIGGRAPH Conference Papers}}.
\newblock


\bibitem[Moon et~al\mbox{.}(2007)]%
        {moon07}
\bibfield{author}{\bibinfo{person}{Jonathan~T. Moon}, \bibinfo{person}{Bruce Walter}, {and} \bibinfo{person}{Stephen~R. Marschner}.} \bibinfo{year}{2007}\natexlab{}.
\newblock \showarticletitle{Rendering Discrete Random Media Using Precomputed Scattering Solutions}. In \bibinfo{booktitle}{\emph{Proceedings of EGSR}}.
\newblock


\bibitem[Moon(1996)]%
        {EM}
\bibfield{author}{\bibinfo{person}{T.K. Moon}.} \bibinfo{year}{1996}\natexlab{}.
\newblock \showarticletitle{The expectation-maximization algorithm}.
\newblock \bibinfo{journal}{\emph{IEEE Signal Processing Magazine}} \bibinfo{volume}{13}, \bibinfo{number}{6} (\bibinfo{year}{1996}).
\newblock


\bibitem[M\"uller et~al\mbox{.}(2022)]%
        {mueller2022instant}
\bibfield{author}{\bibinfo{person}{Thomas M\"uller}, \bibinfo{person}{Alex Evans}, \bibinfo{person}{Christoph Schied}, {and} \bibinfo{person}{Alexander Keller}.} \bibinfo{year}{2022}\natexlab{}.
\newblock \showarticletitle{Instant Neural Graphics Primitives with a Multiresolution Hash Encoding}.
\newblock \bibinfo{journal}{\emph{ACM Trans. Graph.}} \bibinfo{volume}{41}, \bibinfo{number}{4} (\bibinfo{year}{2022}).
\newblock


\bibitem[M\"{u}ller et~al\mbox{.}(2016)]%
        {mueller16efficient}
\bibfield{author}{\bibinfo{person}{Thomas M\"{u}ller}, \bibinfo{person}{Marios Papas}, \bibinfo{person}{Markus Gross}, \bibinfo{person}{Wojciech Jarosz}, {and} \bibinfo{person}{Jan Nov\'{a}k}.} \bibinfo{year}{2016}\natexlab{}.
\newblock \showarticletitle{Efficient Rendering of Heterogeneous Polydisperse Granular Media}.
\newblock \bibinfo{journal}{\emph{ACM Trans. Graph.}} \bibinfo{volume}{35}, \bibinfo{number}{6} (\bibinfo{year}{2016}).
\newblock


\bibitem[Mu{\~n}oz(2014)]%
        {munoz2014higher}
\bibfield{author}{\bibinfo{person}{Adolfo Mu{\~n}oz}.} \bibinfo{year}{2014}\natexlab{}.
\newblock \showarticletitle{Higher order ray marching}.
\newblock \bibinfo{journal}{\emph{Computer Graphics Forum}} \bibinfo{volume}{33}, \bibinfo{number}{8} (\bibinfo{year}{2014}).
\newblock


\bibitem[Museth(2013)]%
        {museth2013vdb}
\bibfield{author}{\bibinfo{person}{Ken Museth}.} \bibinfo{year}{2013}\natexlab{}.
\newblock \showarticletitle{VDB: High-resolution sparse volumes with dynamic topology}.
\newblock \bibinfo{journal}{\emph{ACM Trans. Graph.}} \bibinfo{volume}{32}, \bibinfo{number}{3} (\bibinfo{year}{2013}).
\newblock


\bibitem[Museth(2021)]%
        {museth2021nanovdb}
\bibfield{author}{\bibinfo{person}{Ken Museth}.} \bibinfo{year}{2021}\natexlab{}.
\newblock \showarticletitle{NanoVDB: A GPU-friendly and portable VDB data structure for real-time rendering and simulation}. In \bibinfo{booktitle}{\emph{ACM SIGGRAPH 2021 Talks}}.
\newblock


\bibitem[Neyret(1998)]%
        {neyret98}
\bibfield{author}{\bibinfo{person}{Fabrice Neyret}.} \bibinfo{year}{1998}\natexlab{}.
\newblock \showarticletitle{Modeling, Animating, and Rendering Complex Scenes Using Volumetric Textures}.
\newblock \bibinfo{journal}{\emph{{IEEE} Trans. Vis. Comput. Graph.}} \bibinfo{volume}{4}, \bibinfo{number}{1} (\bibinfo{year}{1998}).
\newblock


\bibitem[Nimier-David et~al\mbox{.}(2022)]%
        {nimierdavid2022unbiased}
\bibfield{author}{\bibinfo{person}{Merlin Nimier-David}, \bibinfo{person}{Thomas M\"uller}, \bibinfo{person}{Alexander Keller}, {and} \bibinfo{person}{Wenzel Jakob}.} \bibinfo{year}{2022}\natexlab{}.
\newblock \showarticletitle{Unbiased Inverse Volume Rendering with Differential Trackers}.
\newblock \bibinfo{journal}{\emph{ACM Trans. Graph.}} \bibinfo{volume}{41}, \bibinfo{number}{4} (\bibinfo{year}{2022}).
\newblock


\bibitem[Nindel et~al\mbox{.}(2021)]%
        {nindel2021gradient}
\bibfield{author}{\bibinfo{person}{Thomas~Klaus Nindel}, \bibinfo{person}{Tom{\'a}{\v{s}} Iser}, \bibinfo{person}{Tobias Rittig}, \bibinfo{person}{Alexander Wilkie}, {and} \bibinfo{person}{Jaroslav K{\v{r}}iv{\'a}nek}.} \bibinfo{year}{2021}\natexlab{}.
\newblock \showarticletitle{A gradient-based framework for 3D print appearance optimization}.
\newblock \bibinfo{journal}{\emph{ACM Trans. Graph.}} \bibinfo{volume}{40}, \bibinfo{number}{4} (\bibinfo{year}{2021}).
\newblock


\bibitem[Nov{\'a}k et~al\mbox{.}(2018)]%
        {novak2018monte}
\bibfield{author}{\bibinfo{person}{Jan Nov{\'a}k}, \bibinfo{person}{Iliyan Georgiev}, \bibinfo{person}{Johannes Hanika}, {and} \bibinfo{person}{Wojciech Jarosz}.} \bibinfo{year}{2018}\natexlab{}.
\newblock \showarticletitle{Monte Carlo methods for volumetric light transport simulation}.
\newblock \bibinfo{journal}{\emph{Computer Graphics Forum}} \bibinfo{volume}{37}, \bibinfo{number}{2} (\bibinfo{year}{2018}).
\newblock


\bibitem[Nov\'{a}k et~al\mbox{.}(2014)]%
        {novak14}
\bibfield{author}{\bibinfo{person}{Jan Nov\'{a}k}, \bibinfo{person}{Andrew Selle}, {and} \bibinfo{person}{Wojciech Jarosz}.} \bibinfo{year}{2014}\natexlab{}.
\newblock \showarticletitle{Residual ratio tracking for estimating attenuation in participating media}.
\newblock \bibinfo{journal}{\emph{ACM Trans. Graph.}} \bibinfo{volume}{33}, \bibinfo{number}{6} (\bibinfo{year}{2014}).
\newblock


\bibitem[Parker et~al\mbox{.}(2010)]%
        {optix}
\bibfield{author}{\bibinfo{person}{Steven~G. Parker}, \bibinfo{person}{James Bigler}, \bibinfo{person}{Andreas Dietrich}, \bibinfo{person}{Heiko Friedrich}, \bibinfo{person}{Jared Hoberock}, \bibinfo{person}{David Luebke}, \bibinfo{person}{David McAllister}, \bibinfo{person}{Morgan McGuire}, \bibinfo{person}{Keith Morley}, \bibinfo{person}{Austin Robison}, {and} \bibinfo{person}{Martin Stich}.} \bibinfo{year}{2010}\natexlab{}.
\newblock \showarticletitle{OptiX: A General Purpose Ray Tracing Engine}.
\newblock \bibinfo{journal}{\emph{ACM Trans. Graph.}} \bibinfo{volume}{29}, \bibinfo{number}{4} (\bibinfo{year}{2010}).
\newblock


\bibitem[Perlin and Hoffert(1989)]%
        {kenperlin89}
\bibfield{author}{\bibinfo{person}{K. Perlin} {and} \bibinfo{person}{E.~M. Hoffert}.} \bibinfo{year}{1989}\natexlab{}.
\newblock \showarticletitle{Hypertexture}. In \bibinfo{booktitle}{\emph{Proceedings of SIGGRAPH}}. \bibinfo{numpages}{10}~pages.
\newblock


\bibitem[Pharr et~al\mbox{.}(2023)]%
        {pbrt}
\bibfield{author}{\bibinfo{person}{Matt Pharr}, \bibinfo{person}{Wenzel Jakob}, {and} \bibinfo{person}{Greg Humphreys}.} \bibinfo{year}{2023}\natexlab{}.
\newblock \bibinfo{booktitle}{\emph{Physically based rendering: From theory to implementation}}.
\newblock \bibinfo{publisher}{MIT Press}.
\newblock


\bibitem[Raab et~al\mbox{.}(2006)]%
        {raab2006unbiased}
\bibfield{author}{\bibinfo{person}{Matthias Raab}, \bibinfo{person}{Daniel Seibert}, {and} \bibinfo{person}{Alexander Keller}.} \bibinfo{year}{2006}\natexlab{}.
\newblock \showarticletitle{Unbiased global illumination with participating media}.
\newblock In \bibinfo{booktitle}{\emph{Monte Carlo and Quasi-Monte Carlo Methods 2006}}.
\newblock


\bibitem[Reiser et~al\mbox{.}(2023)]%
        {reiser2023merf}
\bibfield{author}{\bibinfo{person}{Christian Reiser}, \bibinfo{person}{Rick Szeliski}, \bibinfo{person}{Dor Verbin}, \bibinfo{person}{Pratul Srinivasan}, \bibinfo{person}{Ben Mildenhall}, \bibinfo{person}{Andreas Geiger}, \bibinfo{person}{Jon Barron}, {and} \bibinfo{person}{Peter Hedman}.} \bibinfo{year}{2023}\natexlab{}.
\newblock \showarticletitle{Merf: Memory-efficient radiance fields for real-time view synthesis in unbounded scenes}.
\newblock \bibinfo{journal}{\emph{ACM Trans. Graph.}} \bibinfo{volume}{42}, \bibinfo{number}{4} (\bibinfo{year}{2023}).
\newblock


\bibitem[Sainz and Pajarola(2004)]%
        {sainz04}
\bibfield{author}{\bibinfo{person}{Miguel Sainz} {and} \bibinfo{person}{Renato Pajarola}.} \bibinfo{year}{2004}\natexlab{}.
\newblock \showarticletitle{Point-based rendering techniques}.
\newblock \bibinfo{journal}{\emph{Comput. Graph.}} \bibinfo{volume}{28}, \bibinfo{number}{6} (\bibinfo{year}{2004}).
\newblock


\bibitem[Schröder et~al\mbox{.}(2011)]%
        {shroder11}
\bibfield{author}{\bibinfo{person}{Kai Schröder}, \bibinfo{person}{Reinhard Klein}, {and} \bibinfo{person}{Arno Zinke}.} \bibinfo{year}{2011}\natexlab{}.
\newblock \showarticletitle{A Volumetric Approach to Predictive Rendering of Fabrics}.
\newblock \bibinfo{journal}{\emph{Computer Graphics Forum}} \bibinfo{volume}{30}, \bibinfo{number}{4} (\bibinfo{year}{2011}).
\newblock


\bibitem[Spanier(1966)]%
        {spanier1966two}
\bibfield{author}{\bibinfo{person}{Jerome Spanier}.} \bibinfo{year}{1966}\natexlab{}.
\newblock \showarticletitle{Two pairs of families of estimators for transport problems}.
\newblock \bibinfo{journal}{\emph{SIAM J. Appl. Math.}} \bibinfo{volume}{14}, \bibinfo{number}{4} (\bibinfo{year}{1966}).
\newblock


\bibitem[Spanier and Gelbard(2008)]%
        {spanier2008monte}
\bibfield{author}{\bibinfo{person}{Jerome Spanier} {and} \bibinfo{person}{Ely~M Gelbard}.} \bibinfo{year}{2008}\natexlab{}.
\newblock \bibinfo{booktitle}{\emph{Monte Carlo principles and neutron transport problems}}.
\newblock \bibinfo{publisher}{Courier Corporation}.
\newblock


\bibitem[Sumin et~al\mbox{.}(2019)]%
        {sumin19}
\bibfield{author}{\bibinfo{person}{Denis Sumin}, \bibinfo{person}{Tobias Rittig}, \bibinfo{person}{Vahid Babaei}, \bibinfo{person}{Tim Weyrich}, \bibinfo{person}{Thomas Nindel}, \bibinfo{person}{Piotr Didyk}, \bibinfo{person}{Bernd Bickel}, \bibinfo{person}{Jaroslav K{\v{r}}iv{\'{a}}nek}, \bibinfo{person}{Alexander Wilkie}, {and} \bibinfo{person}{Karol Myszkowski}.} \bibinfo{year}{2019}\natexlab{}.
\newblock \showarticletitle{Geometry-Aware Scattering Compensation for 3D Printing}.
\newblock \bibinfo{journal}{\emph{ACM Trans. Graph.}} \bibinfo{volume}{38}, \bibinfo{number}{4} (\bibinfo{year}{2019}).
\newblock


\bibitem[Sutton et~al\mbox{.}(1999)]%
        {stutton99regular}
\bibfield{author}{\bibinfo{person}{T~M Sutton}, \bibinfo{person}{F~B Brown}, \bibinfo{person}{F~G Bischoff}, \bibinfo{person}{D~B MacMillan}, \bibinfo{person}{C~L Ellis}, \bibinfo{person}{J~T Ward}, \bibinfo{person}{C~T Ballinger}, \bibinfo{person}{D~J Kelly}, {and} \bibinfo{person}{L Schindler}.} \bibinfo{year}{1999}\natexlab{}.
\newblock \showarticletitle{The Physical Models and Statistical Procedures Used in the RACER Monte Carlo Code}.
\newblock  (\bibinfo{year}{1999}).
\newblock
\urldef\tempurl%
\url{https://doi.org/10.2172/767449}
\showDOI{\tempurl}


\bibitem[Szirmay-Kalos et~al\mbox{.}(2017)]%
        {szirmay2017unbiased}
\bibfield{author}{\bibinfo{person}{L{\'a}szl{\'o} Szirmay-Kalos}, \bibinfo{person}{Iliyan Georgiev}, \bibinfo{person}{Mil{\'a}n Magdics}, \bibinfo{person}{Bal{\'a}zs Moln{\'a}r}, {and} \bibinfo{person}{D{\'a}vid L{\'e}gr{\'a}dy}.} \bibinfo{year}{2017}\natexlab{}.
\newblock \showarticletitle{Unbiased light transport estimators for inhomogeneous participating media}.
\newblock \bibinfo{journal}{\emph{Computer Graphics Forum}} \bibinfo{volume}{36}, \bibinfo{number}{2} (\bibinfo{year}{2017}).
\newblock


\bibitem[Szirmay-Kalos et~al\mbox{.}(2011)]%
        {szirmay2011free}
\bibfield{author}{\bibinfo{person}{L{\'a}szl{\'o} Szirmay-Kalos}, \bibinfo{person}{Bal{\'a}zs T{\'o}th}, {and} \bibinfo{person}{Mil{\'a}n Magdics}.} \bibinfo{year}{2011}\natexlab{}.
\newblock \showarticletitle{Free path sampling in high resolution inhomogeneous participating media}.
\newblock \bibinfo{journal}{\emph{Computer Graphics Forum}} \bibinfo{volume}{30}, \bibinfo{number}{1} (\bibinfo{year}{2011}).
\newblock


\bibitem[Tuy and Tuy(1984)]%
        {tuy1984direct}
\bibfield{author}{\bibinfo{person}{Heang~K Tuy} {and} \bibinfo{person}{Lee~Tan Tuy}.} \bibinfo{year}{1984}\natexlab{}.
\newblock \showarticletitle{Direct 2-D display of 3-D objects}.
\newblock \bibinfo{journal}{\emph{IEEE Computer Graphics and Applications}} \bibinfo{volume}{4}, \bibinfo{number}{10} (\bibinfo{year}{1984}).
\newblock


\bibitem[Vicini et~al\mbox{.}(2021a)]%
        {vicini2021non}
\bibfield{author}{\bibinfo{person}{Delio Vicini}, \bibinfo{person}{Wenzel Jakob}, {and} \bibinfo{person}{Anton Kaplanyan}.} \bibinfo{year}{2021}\natexlab{a}.
\newblock \showarticletitle{A non-exponential transmittance model for volumetric scene representations}.
\newblock \bibinfo{journal}{\emph{ACM Trans. Graph.}} \bibinfo{volume}{40}, \bibinfo{number}{4} (\bibinfo{year}{2021}).
\newblock


\bibitem[Vicini et~al\mbox{.}(2021b)]%
        {Vicini2021PathReplay}
\bibfield{author}{\bibinfo{person}{Delio Vicini}, \bibinfo{person}{Sébastien Speierer}, {and} \bibinfo{person}{Wenzel Jakob}.} \bibinfo{year}{2021}\natexlab{b}.
\newblock \showarticletitle{Path Replay Backpropagation: Differentiating Light Paths using Constant Memory and Linear Time}.
\newblock \bibinfo{journal}{\emph{ACM Trans. Graph.}} \bibinfo{volume}{40}, \bibinfo{number}{4} (\bibinfo{year}{2021}).
\newblock


\bibitem[Vorba et~al\mbox{.}(2014)]%
        {vorba2014line}
\bibfield{author}{\bibinfo{person}{Ji{\v{r}}{\'\i} Vorba}, \bibinfo{person}{Ond{\v{r}}ej Karl{\'\i}k}, \bibinfo{person}{Martin {\v{S}}ik}, \bibinfo{person}{Tobias Ritschel}, {and} \bibinfo{person}{Jaroslav K{\v{r}}iv{\'a}nek}.} \bibinfo{year}{2014}\natexlab{}.
\newblock \showarticletitle{On-line learning of parametric mixture models for light transport simulation}.
\newblock \bibinfo{journal}{\emph{ACM Trans. Graph.}} \bibinfo{volume}{33}, \bibinfo{number}{4} (\bibinfo{year}{2014}).
\newblock


\bibitem[Wang et~al\mbox{.}(2004)]%
        {ssim}
\bibfield{author}{\bibinfo{person}{Zhou Wang}, \bibinfo{person}{A.C. Bovik}, \bibinfo{person}{H.R. Sheikh}, {and} \bibinfo{person}{E.P. Simoncelli}.} \bibinfo{year}{2004}\natexlab{}.
\newblock \showarticletitle{Image quality assessment: from error visibility to structural similarity}.
\newblock \bibinfo{journal}{\emph{IEEE Transactions on Image Processing}} \bibinfo{volume}{13}, \bibinfo{number}{4} (\bibinfo{year}{2004}).
\newblock


\bibitem[Woodcock(1965)]%
        {woodcock}
\bibfield{author}{\bibinfo{person}{E Woodcock}.} \bibinfo{year}{1965}\natexlab{}.
\newblock \showarticletitle{Techniques used in the GEM code for Monte Carlo neutronics calculations in reactors and other systems of complex geometry}. In \bibinfo{booktitle}{\emph{Proceedings of the Conference on Applications of Computing Methods to Reactor Problems}}, Vol.~\bibinfo{volume}{557}.
\newblock


\bibitem[Xu et~al\mbox{.}(2013)]%
        {Xu13sigasia}
\bibfield{author}{\bibinfo{person}{Kun Xu}, \bibinfo{person}{Wei-Lun Sun}, \bibinfo{person}{Zhao Dong}, \bibinfo{person}{Dan-Yong Zhao}, \bibinfo{person}{Run-Dong Wu}, {and} \bibinfo{person}{Shi-Min Hu}.} \bibinfo{year}{2013}\natexlab{}.
\newblock \showarticletitle{Anisotropic Spherical Gaussians}.
\newblock \bibinfo{journal}{\emph{ACM Trans. Graph.}} \bibinfo{volume}{32}, \bibinfo{number}{6} (\bibinfo{year}{2013}).
\newblock


\bibitem[Xu et~al\mbox{.}(2022)]%
        {pointnerf}
\bibfield{author}{\bibinfo{person}{Qiangeng Xu}, \bibinfo{person}{Zexiang Xu}, \bibinfo{person}{Julien Philip}, \bibinfo{person}{Sai Bi}, \bibinfo{person}{Zhixin Shu}, \bibinfo{person}{Kalyan Sunkavalli}, {and} \bibinfo{person}{Ulrich Neumann}.} \bibinfo{year}{2022}\natexlab{}.
\newblock \showarticletitle{Point-nerf: Point-based neural radiance fields}. In \bibinfo{booktitle}{\emph{Proceedings of CVPR}}.
\newblock


\bibitem[Yan et~al\mbox{.}(2016)]%
        {yan2016position}
\bibfield{author}{\bibinfo{person}{Ling-Qi Yan}, \bibinfo{person}{Milo{\v{s}} Ha{\v{s}}an}, \bibinfo{person}{Steve Marschner}, {and} \bibinfo{person}{Ravi Ramamoorthi}.} \bibinfo{year}{2016}\natexlab{}.
\newblock \showarticletitle{Position-normal distributions for efficient rendering of specular microstructure}.
\newblock \bibinfo{journal}{\emph{ACM Trans. Graph.}} \bibinfo{volume}{35}, \bibinfo{number}{4} (\bibinfo{year}{2016}).
\newblock


\bibitem[Yu et~al\mbox{.}(2021)]%
        {yu2021plenoctrees}
\bibfield{author}{\bibinfo{person}{Alex Yu}, \bibinfo{person}{Ruilong Li}, \bibinfo{person}{Matthew Tancik}, \bibinfo{person}{Hao Li}, \bibinfo{person}{Ren Ng}, {and} \bibinfo{person}{Angjoo Kanazawa}.} \bibinfo{year}{2021}\natexlab{}.
\newblock \showarticletitle{{PlenOctrees} for Real-time Rendering of Neural Radiance Fields}. In \bibinfo{booktitle}{\emph{Proceedings of ICCV}}.
\newblock


\bibitem[Yue et~al\mbox{.}(2010)]%
        {yue2010unbiased}
\bibfield{author}{\bibinfo{person}{Yonghao Yue}, \bibinfo{person}{Kei Iwasaki}, \bibinfo{person}{Bing-Yu Chen}, \bibinfo{person}{Yoshinori Dobashi}, {and} \bibinfo{person}{Tomoyuki Nishita}.} \bibinfo{year}{2010}\natexlab{}.
\newblock \showarticletitle{Unbiased, adaptive stochastic sampling for rendering inhomogeneous participating media}.
\newblock \bibinfo{journal}{\emph{ACM Trans. Graph.}} \bibinfo{volume}{29}, \bibinfo{number}{6} (\bibinfo{year}{2010}).
\newblock


\bibitem[Zhang et~al\mbox{.}(2019)]%
        {Zhang2019drt}
\bibfield{author}{\bibinfo{person}{Cheng Zhang}, \bibinfo{person}{Lifan Wu}, \bibinfo{person}{Changxi Zheng}, \bibinfo{person}{Ioannis Gkioulekas}, \bibinfo{person}{Ravi Ramamoorthi}, {and} \bibinfo{person}{Shuang Zhao}.} \bibinfo{year}{2019}\natexlab{}.
\newblock \showarticletitle{A Differential Theory of Radiative Transfer}.
\newblock \bibinfo{journal}{\emph{ACM Trans. Graph.}} \bibinfo{volume}{38}, \bibinfo{number}{6} (\bibinfo{year}{2019}).
\newblock


\bibitem[Zhang et~al\mbox{.}(2020)]%
        {zhang2020nerf++}
\bibfield{author}{\bibinfo{person}{Kai Zhang}, \bibinfo{person}{Gernot Riegler}, \bibinfo{person}{Noah Snavely}, {and} \bibinfo{person}{Vladlen Koltun}.} \bibinfo{year}{2020}\natexlab{}.
\newblock \showarticletitle{Nerf++: Analyzing and improving neural radiance fields}.
\newblock \bibinfo{journal}{\emph{arXiv preprint arXiv:2010.07492}} (\bibinfo{year}{2020}).
\newblock


\bibitem[Zhao et~al\mbox{.}(2011)]%
        {zhao11}
\bibfield{author}{\bibinfo{person}{Shuang Zhao}, \bibinfo{person}{Wenzel Jakob}, \bibinfo{person}{Steve Marschner}, {and} \bibinfo{person}{Kavita Bala}.} \bibinfo{year}{2011}\natexlab{}.
\newblock \showarticletitle{Building Volumetric Appearance Models of Fabric Using Micro CT Imaging}.
\newblock \bibinfo{journal}{\emph{ACM Trans. Graph.}} \bibinfo{volume}{30}, \bibinfo{number}{4} (\bibinfo{year}{2011}).
\newblock


\bibitem[Zhao et~al\mbox{.}(2016)]%
        {zhao2016downsample}
\bibfield{author}{\bibinfo{person}{Shaung Zhao}, \bibinfo{person}{Lifan Wu}, \bibinfo{person}{Fr\'edo Durand}, {and} \bibinfo{person}{Ravi Ramamoorthi}.} \bibinfo{year}{2016}\natexlab{}.
\newblock \showarticletitle{Downsampling Scattering Parameters for Rendering Anisotropic Media}.
\newblock \bibinfo{journal}{\emph{ACM Trans. Graph.}} \bibinfo{volume}{35}, \bibinfo{number}{6} (\bibinfo{year}{2016}).
\newblock


\bibitem[Zhu et~al\mbox{.}(2021)]%
        {zhu2021neural}
\bibfield{author}{\bibinfo{person}{Junqiu Zhu}, \bibinfo{person}{Yaoyi Bai}, \bibinfo{person}{Zilin Xu}, \bibinfo{person}{Steve Bako}, \bibinfo{person}{Edgar Vel{\'a}zquez-Armend{\'a}riz}, \bibinfo{person}{Lu Wang}, \bibinfo{person}{Pradeep Sen}, \bibinfo{person}{Milo{\v{s}} Ha{\v{s}}an}, {and} \bibinfo{person}{Ling-Qi Yan}.} \bibinfo{year}{2021}\natexlab{}.
\newblock \showarticletitle{Neural complex luminaires: representation and rendering}.
\newblock \bibinfo{journal}{\emph{ACM Trans. Graph.}} \bibinfo{volume}{40}, \bibinfo{number}{4} (\bibinfo{year}{2021}).
\newblock


\bibitem[Zwicker et~al\mbox{.}(2001)]%
        {ewasplattingzwicker04}
\bibfield{author}{\bibinfo{person}{M. Zwicker}, \bibinfo{person}{H. Pfister}, \bibinfo{person}{J. van Baar}, {and} \bibinfo{person}{M. Gross}.} \bibinfo{year}{2001}\natexlab{}.
\newblock \showarticletitle{EWA volume splatting}. In \bibinfo{booktitle}{\emph{Proceedings of Visualization}}.
\newblock


\end{thebibliography}
\appendix
\section{Single-scattering Albedo and Phase Function}
\label{app:opt_params}
For completeness, here we extend the mixture-based optical properties' definition introduced in Section~\ref{sec:method} to kernel-mixture-based single-scattering albedo and phase function. They are computed similarly to the extinction coefficient in Equation~\eqref{eq:extinction}, with the key difference of requiring normalization. Thus, for a set of $N$ primitives, the single-scattering albedo $\sAlbedo(\px)$ and phase function $\sPF(\px,\wi\rarrow\wo)$ are defined as%
\begin{align}
    \sAlbedo(\px) &= \sum_{i=1}^N \frac{\sCross_i \sKernel_i(\px) \sAlbedo{}_i}{\sum_{i=1}^N \sCross_i\sKernel_i(\px)}, \\ 
    \sPF(\px,\wi\rarrow\wo) &= \sum_{i=1}^N \frac{\sCross_i\sKernel_i(\px)\,\sAlbedo{}_i\,\sPF{}_{,i}(\wi\rarrow\wo)}{\sAlbedo(\px)},
\end{align}%
respectively, with $\sAlbedo{}_i$ and $\sPF{}_{,i}(\wi\rarrow\wo)$ the single-scattering albedo and phase function of primitive $\sPrim_i$.

\section{Kernel Transmittance Expressions}
\label{app:kernel_transmittance}

\subsection{Gaussian kernel}
Assuming our ray direction normalized and whitened (linearly transformed to the local space of the Gaussian shell's ellipsoid, where it is a linear sphere, we can compute the cumulative distribution function (CDF) of a single weighted 3D Gaussian primitive along the given ray as

\begin{align}
\sOptDepth_i(\px_{-t_s},\px_{t_s}) = \sCross_i \int_{-t_s}^{t_s} g_i(\px_t) \,\diff{t} = G
    e^{ -\frac{1}{2} \left(a - b^2 \right)} 
    \left[\operatorname{erf} \left( t_s\sqrt{\frac{1}{2}} \right) \right]
\end{align}

with

\begin{equation}
    \begin{aligned}
        a = \px_{0} \times \px_{0} , \quad
        b = \px_{0} \times \dir , \quad
        G = \frac{\sCross_i}{2\pi\sqrt{v^T\Sigma_i^{-1}v|\Sigma_i|}}
    \end{aligned}
\end{equation}

This equation will be plugged in Equations~\eqref{eq:rec_transmittance} and \eqref{eq:primi_optdepth} to compute the contribution of each overlapping primitive to the accumulated transmittance along the ray. However, to simplify this equation and accelerate rendering, we can resort to integrating along the whole domain of the ray $t_0 = -\infty, t_1 = \infty$ when no overlaps are detected between the integration limits (shell) of a given primitive. This further simplifies the integral to
\begin{align}
\label{eq:full_integral}
    \sOptDepth_i(\px_{-\infty},\px_{\infty}) = G
    e^{ -\frac{1}{2} \left(a - b^2 \right)}
\end{align}

Epanechnikov kernel
\revision{Again, assuming our ray direction normalized, we can compute the CDF of a single 3D Epanechnikov primitive along the given ray as}

\begin{align}
\label{eq:epa_kernel_supp}
    \sOptDepth_i(\px_{-t_s},\px_{t_s}) =\frac{15}{8\pi(7^3\left | \Sigma  \right |)^\frac{1}{2}}\int_{t_0}^{t_1} \nonumber &\\
    \left[1 - \frac{1}{7}\left(\frac{(x_{0,x}+\omega_xt)^2}{S_x^2}+\frac{(x_{0,y}+\omega_yt)^2}{S_y^2}+\frac{(x_{0,z}+\omega_zt)^2}{S_z^2}\right)\right] \diff{t} = \nonumber\\
    K_{\text{norm}}(\mathcal{K}_3{}(t_0^3 - t_1^3)+\mathcal{K}_2{}(t_0^2 - t_1^2) +\mathcal{K}_1{}(t_0 - t_1)),
\end{align}
with
\begin{gather*}
\label{eq:segment_integral_consts_epa}
\mathcal{K}_1 = 3((x_{0,x}^2-7S_x^2)S_y^2+x_{0,y}^2S_x^2)S_z^2+x_{0,z}^2S_x^2S_y^2),\\
\mathcal{K}_2 = 3(x_{0,z}S_x^2S_y^2\omega_z+x_{0,y}S_x^2S_z^2\omega_y+x_{0,x}S_y^2S_z^2\omega_x),\\
\mathcal{K}_3 = S_x^2S_y^2\omega_z^2+S_x^2S_z^2\omega_y^2+S_y^2S_z^2\omega_x^2\\
\mathcal{K}_{\text{norm}} =\frac{15}{168\pi7^\frac{3}{2} \left | \Sigma  \right |^2} ,
\end{gather*}
and $t_0, t_1$  defining the segment along the ray being integrated.

\section{Integrator Adjoints}
\label{sec:supp_adjoint}
Here we include the adjoints for both the VPPT and VPRF integrators. For simplicity, in the following formulas we omit the set of kernel primitives for all summations and products, as well as the positional dependence.

\paragraph{Adjoint VPPT}
The derivative of Equation~\eqref{eq:vppt} is
\begin{align*}
\delta L
    &= \delta (\beta) \cdot \sAlbedo_S \cdot \sTrans_S \cdot \mu_S \cdot L_i 
    + \beta \cdot \delta (\sAlbedo_S) \cdot \sTrans_S \cdot \mu_S \cdot L_i \\
    &+ \beta \cdot \sAlbedo_S \cdot \delta (\sTrans_S) \cdot \mu_S \cdot L_i 
    + \beta \cdot \sAlbedo_S \cdot \sTrans_S \cdot \delta (\mu_S) \cdot L_i \\
    &+ \beta \cdot \sAlbedo_S \cdot \sTrans_S \cdot \mu_S \cdot \delta (L_i) \\
    \nonumber
\end{align*}
with 
\begin{align}
\delta \sAlbedo_S &= \sum_i \left( \delta (\sAlbedo_i p_i) \cdot \frac{1}{\sum_j p_j} - \delta(p_i) \cdot \frac{1}{(\sum_j p_j)^2} \cdot \sum_j \sAlbedo_j p_j \right ) \nonumber, \\
\delta T_S & = \sum_i \delta (e^{-\tau_i}) \cdot \frac{(\prod_j e^{-\tau_j})}{e^{\tau_i}}, \nonumber \\
\delta \mu_S &= \sum_i \delta \mu_i \nonumber,
\end{align}
and $p_i=\sCross_i \sKernel_i(\px)$.

\paragraph{Adjoint VPRF} 
The derivative of Equation~\eqref{eq:segments_rte2_main} is
\begin{equation}
\delta L
    = \sum_{k=1}^M \delta (\sTrans_{k-1}) L_k + \sTrans_{k-1} \delta(L_k),  \nonumber
\end{equation}
with for the VPRF integrator
\begin{align}
    \delta \sTrans_{k} & = \sum_{j=0}^{k} \delta(\sTrans_j)\frac{\prod_{l=0}^k \sTrans_l}{\sTrans_j}, \nonumber\\
    \delta{L_k} &= \delta\left(1-\prod_{i \in \sSegmentSet_k} e^{-\sOptDepth_i}\right) \sum_{i \in \sSegmentSet_k} \left(\sRedRadE_i \frac{\sOptDepth_i}{\sum_j \sOptDepth_j}\right) \nonumber\\ 
    & + \left(1-\prod_{i \in \sSegmentSet_k} e^{-\sOptDepth_i}\right) \sum_{i \in \sSegmentSet_k} \delta\left(\sRedRadE_i \frac{\sOptDepth_i}{\sum_j \sOptDepth_j}\right), \nonumber
\end{align}
where
\begin{align}
    \delta\left(1-\prod_{i \in \sSegmentSet_k} e^{-\sOptDepth_i}\right) & =
    \sum_{i \in \sSegmentSet_k} \frac{-\delta(e^{-\sOptDepth_i})}{e^{-\sOptDepth_i}} \prod_{j \in \sSegmentSet_k} \delta(e^{-\sOptDepth_j}), \nonumber\\ 
    \delta\left(\sRedRadE_i \frac{\sOptDepth_i}{\sum_j \sOptDepth_j}\right) & = \delta{\sRedRadE_i} \frac{\sOptDepth_i}{\sum_j \sOptDepth_j} + \sRedRadE_i \frac{\delta(\sOptDepth_i)}{\sum_j \sOptDepth_j} \nonumber\\
    & + \sRedRadE_i \frac{\sOptDepth_i \sum_j \delta(\sOptDepth_j)}{(\sum_j \sOptDepth_j)^2} \nonumber.
\end{align}

\end{document}